\documentclass{aastex}
\usepackage{lscape}
\begin{document} 
\title{Properties of Magneto-Hydrodynamic Waves in the Solar Photosphere obtained with \emph{Hinode}}

\author{D. Fujimura\altaffilmark{1,2}, S. Tsuneta\altaffilmark{2}}

\email{daisuke.fujimura@nao.ac.jp}

\altaffiltext{1}{Department of Earth \& Planetary Science, School of Science, University of Tokyo, Bunkyoku, 113-0033 Tokyo, Japan.}

\altaffiltext{2}{National Astronomical Observatory, 2-21-1, Osawa, Mitaka, 181-8588 Tokyo, Japan}

\begin{abstract}
We report the observations of the magneto-hydrodynamic waves propagating along magnetic flux tubes in the solar photosphere. We identified 20 isolated strong peaks (8 peaks for pores and 12 peaks for inter-granular magnetic structure) in the power spectra of the l.o.s. (line-of-sight) magnetic flux, the l.o.s. velocity, and the intensity for 14 different magnetic concentrations. The observation is performed with the spectro-polarimeter of the Solar Optical Telescope aboard the \emph{Hinode} satellite. The oscillation periods are located in 3$-$6 min for the pores and in 4$-$9 min for the inter-granular magnetic elements. These peaks correspond to the magnetic, the velocity, and the intensity fluctuation in time domain with r.m.s. (root mean square) amplitudes of $4-17$ G ($0.3-1.2$ \%), $0.03-0.12$ km s$^{-1}$, and $0.1-1$\%, respectively. Phase differences between the l.o.s. magnetic flux $(\phi _{B})$, the l.o.s. velocity $(\phi _{v})$, the intensities of the line core $(\phi _{I,\rm core})$, and the continuum intensity $(\phi _{I,\rm cont})$ have striking concentrations at around $-90^{\circ}$ for $\phi _{B}-\phi _{v}$ and $\phi _{v}-\phi _{I,\rm core}$, around $180^{\circ}$ for $\phi _{I,\rm core}-\phi _{B}$, and around $10^{\circ}$ for $\phi _{I,\rm core}-\phi _{I,\rm cont}$. Here, for example, $\phi _{B}-\phi _{v} \sim -90^{\circ}$ means that the velocity leads the magnetic field by a quarter of cycle. The observed phase relation between the magnetic  and the photometric intensity fluctuation would not be consistent with that caused by the opacity effect, if the magnetic field strength decreases with height along the oblique line of sight. We suggest that the observed fluctuations are due to longitudinal (sausage mode) and/or transverse (kink mode) MHD waves. The observed phase relation between the fluctuations in the magnetic flux and the velocity is consistent with the superposition of the ascending wave and the descending wave reflected at chromosphere/corona boundary (standing wave). Even with such reflected waves, the residual upward Poynting flux is estimated to be $2.7 \times 10^{6}$ erg cm$^{-2}$ s$^{-1}$ for a case of the kink wave. Seismology of the magnetic flux tubes is possible to obtain various physical parameters from the observed period and amplitude of the oscillations. 
\end{abstract}


\section{Introduction}
Alfv\'en waves or more generally transverse magneto-hydrodynamic waves would play a key role in coronal heating and solar wind acceleration (e.g. Suzuki \& Inutsuka, 2006). Numerous studies about generation, propagation and dissipation of the Alfv\'en waves have been carried out observationally and theoretically (e.g. Ryutova \& Priest, 1993). Alfv\'en waves would be generated in the high-$\beta$ region of the solar atmosphere. Its precise power spectra is, however, not observationally known. Ascending Alfv\'en waves with wavelength longer than the Alfv\'enic scale height may be reflected back at the chromospheric-coronal boundary (Moore et al.1991; An et al. 1989; Hollweg 1978; Suzuki \& Inutsuka, 2005). It is poorly known how much Alfv\'en-wave flux generated in the photosphere is propagated all the way to the corona through the fanning-out flux tubes. High-quality observations to obtain spectra of magnetic fluctuation is of crucial importance to understand coronal heating and acceleration of fast solar wind. 

Ulrich (1996) made the first critical observations, and reported the detection of the magneto-hydrodynamic oscillations with properties of the Alfv\'en waves. He suggested that the observed phase relation between the magnetic field and the velocity perturbation is consistent with the outgoing Alfv\'en waves. The observing aperture of $20\arcsec \times 20\arcsec$ is, however, very large compared with the spatial scale of the flux tubes along which the Alfv\'en waves propagate. Such a large aperture may make it difficult to identify the weak transverse waves with different frequency and phase, which might become evident in higher resolution observations. Velocity and magnetic field oscillations in the sunspot umbra were detected by Bellot Rubio et al. (2000), Lites et al. (1998), Norton et al. (1999), R\"{u}edi et al. (1998), R\"{u}edi \& Solanki (1999), Balthasar (1999), and Settele et al. (2002). R\"{u}edi et al. (1998) and Bellot Rubio et al. (2000) obtained the phase difference of -90$^{\circ}$ and 90$^{\circ}$ between the fluctuations of the line-of-sight velocity and the magnetic field strength $\phi_{v}-\phi_{B}$, respectively. They suggested that the magnetic field fluctuation is caused by the opacity fluctuations that move upward and downward the region where the spectral line profiles are sensitive to magnetic fields. Norton et al. (2001) obtained the center-to-limb dependence of the phase angle between the magnetic and the velocity fluctuations with the \emph{Michelson Doppler Imager} aboard the \emph{SOHO} satellite. They reported that the phase angle is near  $-$90$^{\circ}$ at the disk center and near 0$^{\circ}$ at the limb, and made an important comment that the Alfv\'en waves be more easily observed at the limb. They suggested that the phase relation reported in the paper is not due to the opacity effect. Khomenko et al. (2003) compared the analytical solution of the MHD equations including gravity, inclination of magnetic field, and effects of nonadiabaticity with the observations reported by Bellot Rubio (2000), and concluded that the detected time variation in field strength could be partly due to magnetoacoustic waves. R\"{u}edi \& Cally (2003) suggested that most of the observed magnetic field oscillations is due to the opacity effect caused by temperature and density fluctuations associated with magnetoacoustic waves.

Recently apparent transverse oscillations, which are clear evidence of the Alfv\'en waves, are detected in prominence (Okamoto et al. 2007), in spicules (de Pontieu et al. 2007, He et al. 2009), and in Ca jet (Nishizuka et al., 2008) with the Solar Optical Telescope (SOT; Tsuneta et al. 2008a; Suematsu et al. 2008; Ichimoto et al. 2008; Shimizu et al. 2008) aboard the \emph{Hinode} satellite (Kosugi et al., 2007). These Alfv\'en waves have enough Poynting flux to potentially heat the corona. We, however, cannot rule out the possibility that these waves are the standing Alfv\'en waves. Transverse oscillations of coronal loops are detected by Taroyan et al. (2008), Mariska et al. (2008), and Van Doorsselaere et al. (2008) using the \emph{EUV Imaging Spectrometer} (Calhane et al., 2007) aboard the \emph{Hinode} satellite as well. Ubiquitous upward Alfv\'en waves in the corona are detected by Tomczyk et al. (2007) using the \emph{Coronal Multi-Channel Polarimeter} without magnetic field information. We stress that the observations of the magnetic field fluctuation with the simultaneous velocity and photometric measurement allows us to identify propagating hydromagnetic waves. 

The literatures so far introduced are mainly concerned with the pure Alfv\'en waves. The magnetic fields in the solar atmosphere have a form of isolated magnetic flux tubes embedded in a nearly field-free fluid. Such flux tubes carry the incompressible torsional Alfv\'en waves, and the linearly polarized Alfv\'en waves can exist only in the uniform media. The flux tubes also carry the kink waves (transverse waves) and the sausage waves (longitudinal waves) (e.g. Stix, 2002) instead of the linearly-polarized Alfv\'en waves. Magnetic tension force of the flux tube is the restoring force in the kink mode  (e.g. Spruit, 1981), and is essentially incompressible. The sausage mode with the azimuthal wave number m = 0, as first defined by Defouw (1976) and discussed e.g. in Roberts and Webb (1978) and Ryutova (1981), is related to a {\it slow} magnetoacoustic mode. In the sausage-shaped perturbed boundary of the flux tube, where the flux-tube area increases, the magnetic field decreases, whereas the plasma pressure increases; vice versa. A fast magneto-acoustic mode propagates across the flux tube, and is not localized radially in the vicinity of the flux tube; we do not regard this as a mode of flux tube oscillations. In this paper, we report a clear detection of magnetic, velocity and photometric oscillations of the magnetic flux tubes with the Spectro-Polarimeter (SP) of SOT. The data is extensively analyzed in terms of both the linearly-polarized kink waves and the slow sausage waves, while we will not discuss the torsional Alfv\'en waves due to our constraint in the analysis as we explain later.  

SOT/SP is ideally suited to detect the magneto-hydrodynamic waves propagated along the flux tubes due to its high spatial and time resolution and its high polarimetric and photometric precision (e.g. Ploner $\&$ Solanki, 1997). SOT/SP obtains two spectra of iron lines (Fe I) with wavelengths of 630.15 nm and 630.25 nm, which are suitable for observing lower photosphere (del Toro Iniesta, 2003). Earlier studies about magnetic fluctuations were done in sunspot umbra, since small-scale flux tube ($\sim 1 \arcsec$) fluctuations might be difficult to detect. The high spatial resolution of \emph{Hinode} ($\sim 0\arcsec .16$) allows us to detect the fluctuations in such small-scale flux tubes. Furthermore, stable observations from space allow us to detect clear intensity fluctuations for the first time, and to obtain the phase relations among the fluctuations in the magnetic flux, the velocity, and the intensity. This allows us to examine the opacity effect more in detail.   

For the detection of weak magnetic fluctuations, we prefer to use the Stokes V signal instead of the Stokes Q or U signal because of its much higher sensitivity to magnetic flux. We, thus, intentionally choose magnetic concentrations located away from the disk center to observe possible fluctuation of the transverse magnetic field in the Stokes V signal and the associated velocity signal. We estimate the magnetic field fluctuation associated with the Alfv\'en waves to be about $\delta B = 10$ G by substituting typical values for the photosphere (magnetic field strength of a flux tube $B_0 =2000$ G, Alfv\'en speed $v_{A}=20$ km s$^{-1}$, and velocity fluctuation $\delta v=0.1$ km s$^{-1}$) to the relation about the Alfv\'en wave $\delta v/v_{A}= \delta B/B_{0}$ (e.g. Priest 1981). A detection limit of the longitudinal and transverse magnetic fields observed by the SOT is known to be  1-5 G and 30-50 G, respectively (Tsuneta et al. 2008a). This exercise demonstrates that the SOT/SP can detect the transverse MHD waves in the Stokes V signal with high signal-to-noise ratio, if such MHD waves are present in the photosphere. 

\section{Observations and Data Analysis}
\subsection{Hinode Observation}
The data used in this paper was taken on 2007 February 3 $-$ 5. The region that we observed was NOAA 10940, which moved from west 25.2 to 49.0 degrees in longitude during the course of the observation. The region consists of pores and magnetic flux concentrations located in the inter-granular lanes (Ishikawa et al., 2007), which we hereafter call Inter-granular Magnetic Structure (IMS). The integration time is 4.8 s, and the field of view is 1\arcsec .92 (EW) by 81\arcsec .92 (NS). The pixel size is 0\arcsec .16. Periodic scanning was done by the SOT/SP for 1 hour or 3 hours depending on the flux tubes with cadence of 67 s. This time resolution allows us to detect magnetohydrodynamic waves with a period longer than 134 s according to the Nyquist criteria.  

We analyzed 14 magnetic flux concentrations as tabulated in Table \ref{table1}. All these magnetic flux concentrations are of positive polarity (magnetic field vector toward the observer). The region \#05 is shown in Fig. \ref{fig1} as an example of the data. The region \#05 contains a pore in a plage region. 

\subsection{Time-Profile Data Analysis}
We use the Stokes I and V profiles of the Fe I 630.25 nm line to derive the line-of-sight velocity, the line-of-sight magnetic flux, and the intensity. The line-of-sight velocity fluctuation $(\delta v_{\rm los})$ is derived by measuring the Stokes V zero cross position $\lambda_{c}$. The Stokes V profiles reflect the motion of the magnetic atmosphere better than the Stokes I profiles, which also contain the information of the non-magnetic atmosphere. The line-of-sight magnetic flux fluctuation ($\delta \Phi_{\rm los}$) is derived with the help of weak field approximation (Landi degl'Innocenti \& Landolfi 2004) rather than relying on the standard Milne-Eddington inversion (e.g. del Toro Iniesta 2003). The Milne-Eddington least-squares fit is performed to the observed Stokes profiles of the Fe I 630.15 nm and Fe I 630.25 nm with 12 parameters, which may be subject to noise that impedes the detection of fluctuation with amplitude as small as $\delta B/B_{0} \sim 0.4 $ \%. In the weak field approximation, the line-of-sight magnetic flux is proportional to the degree of the circular polarization $CP$ defined by:

\begin{eqnarray}
CP \equiv \frac{\int_{\lambda_{c}-d_{1}}^{\lambda_{c}}V(\lambda ){\rm d}\lambda - \int_{\lambda_{c}}^{\lambda_{c}+d_{1}}V(\lambda ){\rm d}\lambda}{I_{\rm cont}}, \label{equa1}
\end{eqnarray}
where $V(\lambda ) $ is the Stokes profile observed with the SOT/SP, $\lambda_{c}$ is the measured zero cross position of the observed Stokes-V profiles as described above, $d_{1}$ is 43.2 pm, and $I_{\rm cont}$ is the  continuum intensity. The observed Stokes I and V profiles for the region \#5 (Table \ref{table1}) are shown in Fig. \ref{fig2} as an example. Since the integration is done with respect to $\lambda_{c}$, and the integration range is wide enough to encompass the entire profiles, the integral should not have any cross-talk with the velocity. Intensity fluctuations in the line core $(\delta I_{\rm core})$ and in the continuum $(\delta I_{\rm cont})$ are derived from the line core intensity $I_{\rm core}$ and continuum intensity $I_{\rm cont}$ defined by,
\begin{eqnarray}
I_{\rm cont} \equiv 4 \Biggl( \int_{\lambda_{c}-d_{2}}^{\lambda_{c}-d_{1}}I(\lambda ){\rm d}\lambda + \int_{\lambda_{c}+d_{1}}^{\lambda_{c}+d_{2}}I(\lambda ){\rm d}\lambda \Biggr) \label{equa2}, \\
I_{\rm core} \equiv 4 \Biggl( \int_{\lambda_{c}-d_{3}}^{\lambda_{c}}I(\lambda ){\rm d}\lambda + \int_{\lambda_{c}}^{\lambda_{c}+d_{3}}I(\lambda ){\rm d}\lambda \Biggr) , \label{equa3}
\end{eqnarray} 
where $I(\lambda )$ is the Stokes I profile observed with the SOT/SP, $d_{2}$ and $d_{3}$ are 54.0 pm and 10.8 pm, respectively, and the factor of 4 is to adjust the difference in the integration range between $CP$ and $I$.

The intrinsic magnetic field strength $(B_0)$ and the filling factor $f$ are derived from the Milne-Eddington inversion to accurately determine the Alfv\'en speed. The intrinsic magnetic field strength $B_0$ is used only for this purpose. The filling factor is defined as the fraction of area occupied with the magnetic field in a pixel (Orozco Su\'arez et al., 2007). The 12 free parameters are intrinsic field strength $(B_0)$, inclination and azimuth for magnetic field vector, line strength, Doppler width, damping factor, Doppler velocity, source function, source gradient, macro turbulence, filling factor (stray-light factor), and the Doppler shift of the stray-light profile.

\begin{figure}
\epsscale{0.8}
\plotone{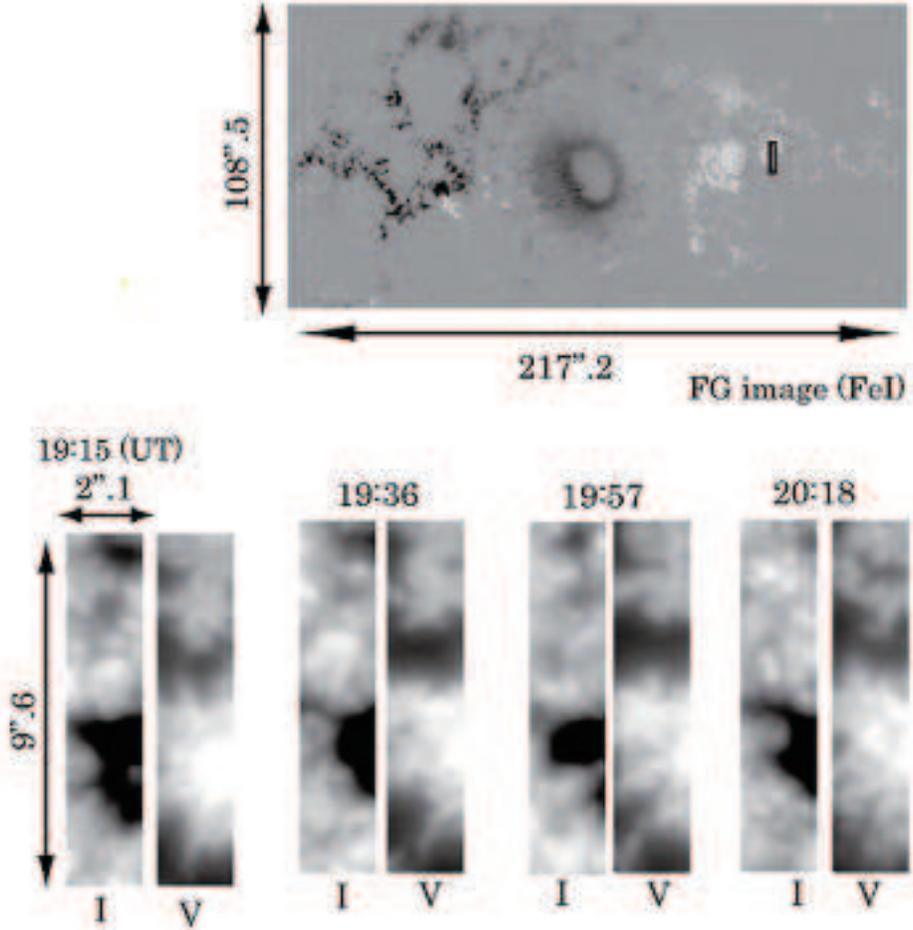}
\caption{
\emph{Top}: The SOT filtergraph (FG) image taken in the FeI 630.25 nm line at 19:45(UT) on 2007 February 3. The field of view is 217\arcsec .1 (EW) $\times$ 108\arcsec .5 (NS). The pixel size is 0\arcsec .108. Exposure time is 90 ms. The black rectangular box indicates the region \#05 (Table \ref{table1}). \emph{Bottom}: Zoomed SP images (Stokes I and V) for the region \#05 taken at 19:15-20:18 (UT) on 2007 February 3. Periodic scanning was done by SP for about 1 hour with a cadence of 67 s. The integration time is 4.8 s. The field of view is 2\arcsec .08 (EW) $\times$ 81\arcsec .92 (NS), part of which is shown here. The pixel size is 0\arcsec .16. The black region in the Stokes I map, which corresponds to the white region in the Stokes V map, is a pore. These images show that the pore lives for at least one hour.    
}
\label{fig1}
\end{figure}

\begin{figure}
\epsscale{0.8}
\plotone{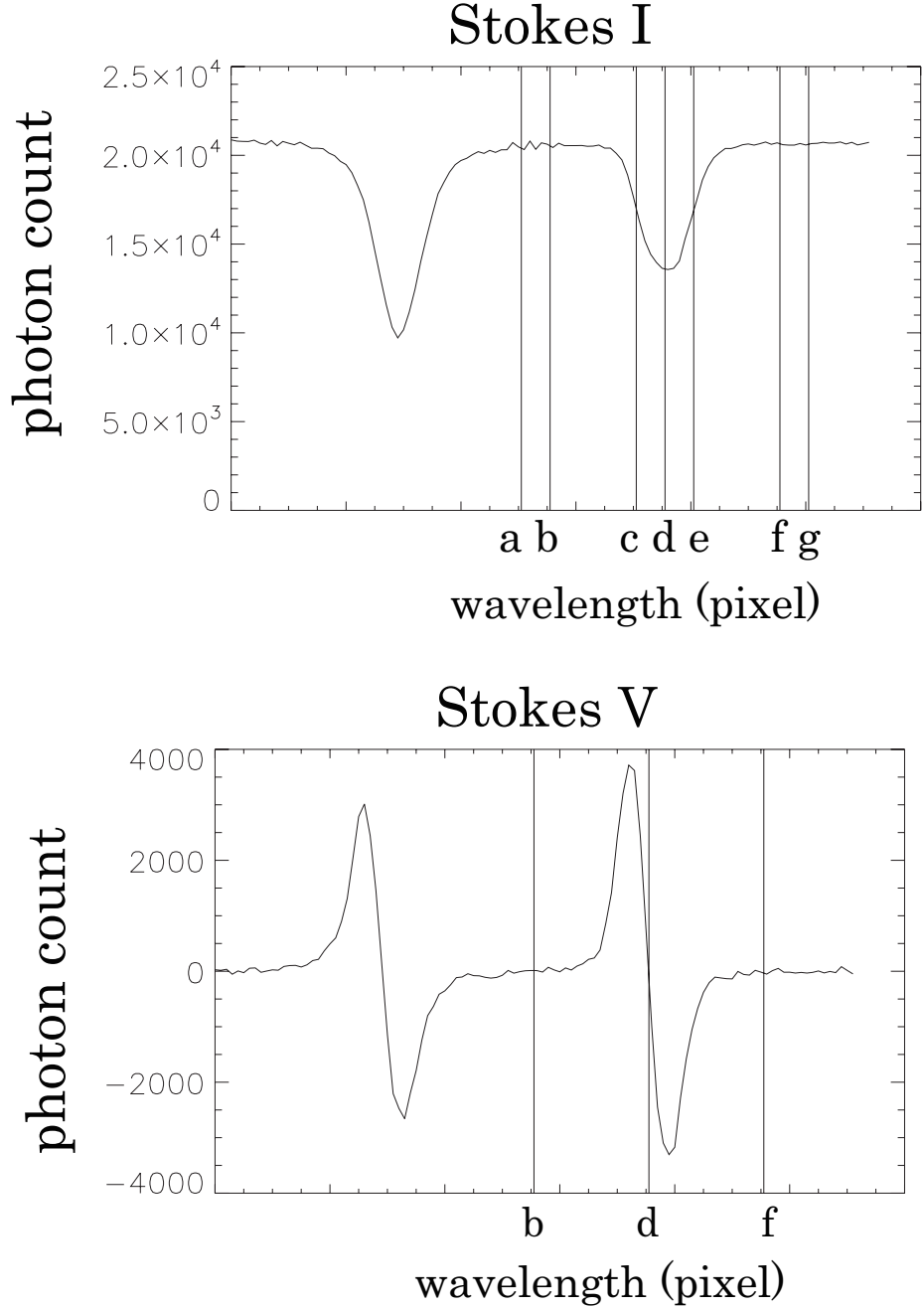}
\caption{
Stokes I profile \emph{(top)} and Stokes V profile for the region \#5 (Table \ref{table1}). Wavelength positions from \emph{a} through \emph{g} define the integration ranges specified by $\lambda_{c}$, $d_{1}$, $d_{2}$, $d_{3}$ in equations (\ref{equa1}) $-$ (\ref{equa3}). \emph{a} through \emph{g} indicate $a:\lambda_{c}-d_{3}$, $b:\lambda_{c}-d_{2}$, $c:\lambda_{c}-d_{1}$, $d:\lambda_{c}$, $e:\lambda_{c}+d_{1}$, $f:\lambda_{c}+d_{2}$, and $g:\lambda_{c}+d_{3}$, respectively.  
}
\label{fig2}
\end{figure}

\begin{figure}
\epsscale{0.8}
\plotone{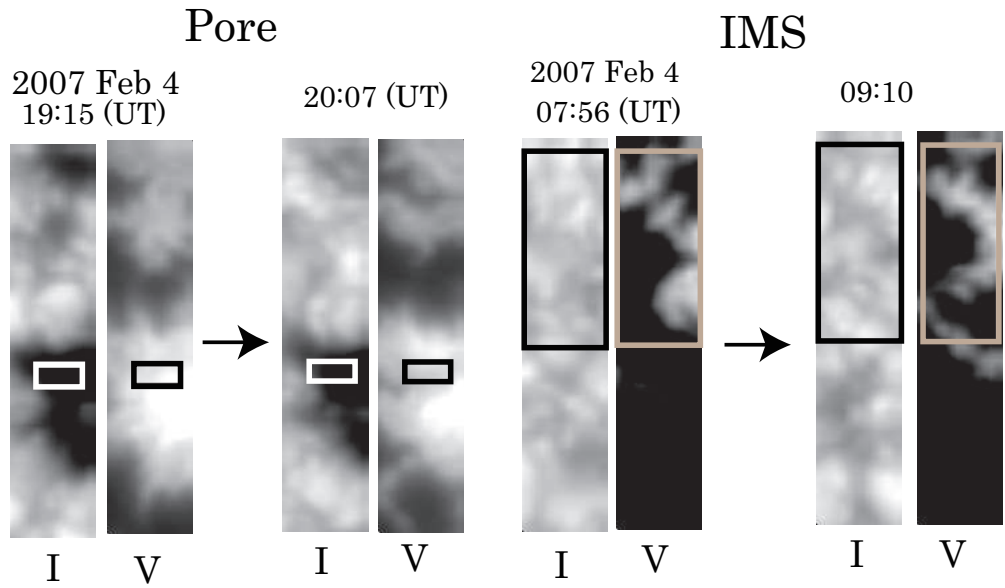}
\caption{
The boxes in each image indicate the region-of-interest (ROI) for the wave analysis. \emph{Left}:
SOT/SP images for the region \#05 (Table \ref{table1}). The ROI with size of 1\arcsec .28 (EW) $\times$ 0\arcsec .64 (NS) is located inside the pore. \emph{Right}: SOT/SP images for the region \#10. The ROI contains inter-granular magnetic Structure (IMS). The images show that the IMS is not maintained for 1 hour. The size of the ROI is 2\arcsec .08 (EW) $\times$ 3\arcsec .52 (NS).
}
\label{fig3}
\end{figure}

We should track the region of interest (ROI), for which the wave analysis is performed, in a Lagrangian way for an extended period of time. In the case of pores, the overall magnetic structure is maintained over 1 hour as shown is Fig. \ref{fig1}. In this case, we set the ROI to cover a portion of a pore. The size of the ROIs for pores is typically 2\arcsec $\times$ 2\arcsec to 2\arcsec $\times$ 5\arcsec. The physical parameters are averaged inside the ROI. On the other hand, IMSs are generally not maintained for 1 hour: magnetic elements may combine, split, or decay within a time period of several tens of minutes. Thus, we set the ROI in this case large enough to encompass the entire magnetic flux concentration in the spatial and temporal domain, and average the physical parameters of the pixels with $CP$ larger than 0.01 inside the ROI. The size of the ROIs for IMSs is typically 1\arcsec $\times$ 1\arcsec to 2\arcsec $\times$ 4\arcsec . Examples of SP images for a pore (ID \#05) and an IMS (ID \#10) are shown in Fig. \ref{fig3}. The average line-of-sight magnetic flux $\overline{\Phi}_{\rm los}$ is given by  
\begin{eqnarray}
\overline{\Phi}_{\rm los}=\frac{ \sum_{i=1}^{N} CP_{i}} {\lambda N} ,\label{equa4}
\end{eqnarray}
where $N$ is the number of pixels inside the ROI for which $CP$ is larger than 0.01. $\lambda$ is the conversion coefficient for converting $CP$ to magnetic flux. $\lambda$ is estimated to be $4.16 \times 10^{-5}$ G$^{-1}$ in the Appendix. Positive values for the line-of-sight magnetic flux and velocity indicate that the are directed towards the observer.

\begin{table} [htbp]
\begin{center}
\tablenotetext{1}{IMS: Inter-granular Magnetic Structure}
\tablenotetext{2}{X-Y coordinate of the target region. X is to the West, and Y is to the North.}
\tablenotetext{3}{Helio-longitudinal angle from the meridional line}
\begin{tabular}{ccccccc} 
\hline
Region & Date & Time & Pore or IMS\tablenotemark{1} & x\tablenotemark{2} & y\tablenotemark{2} & $\theta$\tablenotemark{3} \\ 
ID &   & (UT) &   &  ($\arcsec$) & ($\arcsec$)  & (deg) \\ \hline
\# 01 & 2007 Feb 03 & 13:18 - 14:28 & pore & 410 & 45 & 25 \\ 
\# 02 & 2007 Feb 03 & 14:28 - 15:38 & pore & 410 & -11 & 25 \\ 
\# 03 & 2007 Feb 03 & 12:08 - 13:18 & IMS & 410 & -5 & 25 \\ 
\# 04 & 2007 Feb 03 & 19:15 - 20:27 & pore & 460 & 46 & 29 \\
\# 05 & 2007 Feb 03 & 19:15 - 20:27 & pore & 460 & 41 & 29 \\
\# 06 & 2007 Feb 03 & 19:15 - 20:27 & IMS & 460 & -1 & 29 \\ 
\# 07 & 2007 Feb 03 & 19:15 - 20:27 & pore & 460 & -7 & 29 \\
\# 08 & 2007 Feb 04 & 01:28 - 02:42 & IMS & 510 & 42 & 32 \\ 
\# 09 & 2007 Feb 04 & 01:28 - 02:42 & IMS & 510 & 38 & 32 \\ 
\# 10 & 2007 Feb 04 & 07:56 - 09:10 & IMS & 560 & -21 & 36 \\
\# 11 & 2007 Feb 04 & 14:28 - 15:37 & IMS & 602 & 35 & 39 \\ 
\# 12 & 2007 Feb 04 & 12:45 - 13:54 & IMS & 602 & -5 & 39 \\  
\# 13 & 2007 Feb 04 & 13:31 - 14:40 & IMS & 602 & -27 & 39 \\
\# 14 & 2007 Feb 05 & 06:56 - 08:08 & IMS & 725 & -10 & 49 \\ 
\hline
\end{tabular}
\caption
{
List of observed magnetic flux concentrations. 
} 
\label{table1}
\end{center}
\end{table} 

\section{Power Spectra and Phase Relation}
\begin{figure}
\epsscale{1.0}
\plotone{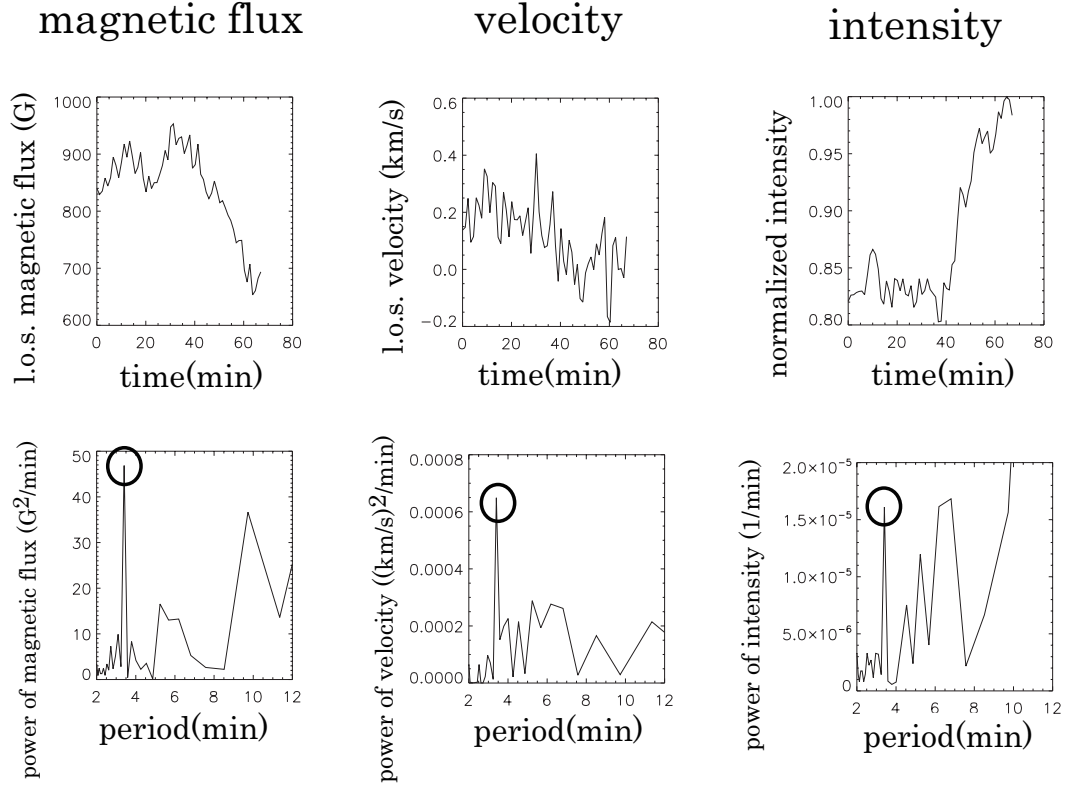}
\caption{
\emph{Top}: Time profiles for the region \#04 of Table \ref{table1}: the line-of-sight (l.o.s.) magnetic flux \emph{(left)}, the l.o.s. velocity \emph{(center)}, and the line core intensity \emph{(right)} as defined by Eq. (\ref{equa3}). The intensity profile is normalized to the peak value of the time profile. Images of the region \#04 are shown in Fig. \ref{fig1}. \emph{Bottom}: The power spectra of the l.o.s. magnetic flux \emph{(left)}, the l.o.s. velocity \emph{(center)}, and the normalized line-core intensity \emph{(right)}. The circles indicate the common, isolated peaks.}
\label{fig4}
\end{figure}

The top panels of Fig. \ref{fig4} show the time profiles of the line-of-sight magnetic flux, the line-of-sight velocity, and the line core intensity for the region \#04 (Table \ref{table1}). We applied the Fourier Transform to all time profiles. The result for the region \#04 is shown in the bottom panels of Fig. \ref{fig4}. The power spectra generally show one or two isolated sharp peaks in the shorter periods, while broader peaks are found in the longer periods, corresponding to a gradual rise and fall in the time profiles. Some of the peaks have the same period in the magnetic and velocity field, and the photometric intensity. We found 20 such common peaks, which are all tabulated in Table \ref{table2}. We analyzed 29 flux tubes, and such common peaks are found in 14 (48\%) flux tubes, which are all tabulated in Table \ref{table1}.  

We derive the r.m.s. amplitudes of the line-of-sight fluctuation in magnetic flux $(\delta \Phi_{\rm los,rms})$ and velocity $(\delta v_{\rm los,rms})$, the line core intensity fluctuations $(\delta I_{\rm core,rms})$, and the continuum intensity fluctuations $(\delta I_{\rm cont,rms})$ at the peak periods in the power spectra. We also obtain phase difference between the fluctuations in the magnetic flux $(\phi _{B})$, the velocity $(\phi _{v})$, the line core intensity $(\phi _{I,\rm core})$, and the continuum intensity $(\phi _{I,\rm cont})$; $\phi _{B}- \phi _{v}$, $\phi _{v}- \phi _{I,\rm core}$, $\phi _{I,\rm core}- \phi _{B}$, and $\phi _{I,\rm core}- \phi _{I,\rm cont} $, all for the peak periods. The phase relations between the fluctuations in the magnetic flux, the velocity, and the intensity fluctuations are of critical importance to identify modes and properties of magneto-hydrodynamic waves as we will see later. 

When $x_n$ is the raw time profile $(0 \leq n \leq N-1)$ ($N$ is the number of data points), then the complex amplitude $X_k$ at the frequency $k$ in the frequency domain is converted to the r.m.s. (root mean square) value of the wave amplitude $A_{k,rms}$ and the phase $\theta_k$ as follows:  
\begin{eqnarray}
X_k =\frac{1}{N} \sum_{n=0}^{N-1} x_n  \exp (-\frac{2 \pi ikn}{N}) ,\label{equa5} \\
A_{k,\rm rms}=\sqrt{2}|X_k|, \label{equa6} \\
\theta _k= \arctan [\frac{\rm Im(X_k )}{\rm Re(X_k )}] .\label{equa7}
\end{eqnarray}
We calculate these values for all the peaks, and Table \ref{table2} lists the l.o.s. magnetic flux $\Phi_{0,\rm los}=B_{0,\rm los}\overline{f} $, where $B_{0,\rm los}$ is the line-of-sight magnetic field and $\overline{f}$ the average filling factor, both of which are derived from Milne-Eddington inversion, the r.m.s. line-of-sight magnetic flux fluctuations $(\delta \Phi_{\rm los, rms})$, $\frac{\delta \Phi_{\rm los,rms}}{\Phi_{\rm 0,los}}$, the r.m.s. line-of-sight velocity fluctuations $(\delta v_{\rm los, rms})$, the r.m.s. line core and continuum intensity fluctuations normalized by the average intensity, $\frac{\delta I_{\rm core,rms}}{\overline{I_{\rm core}}}$ and $\frac{\delta I_{\rm cont,rms}} {\overline{I_{\rm cont}}}$, and the phase difference among magnetic, velocity, and intensity fluctuations; $\phi _{B}-\phi _{v}$, $\phi _{v}-\phi _{I,\rm core}$, $\phi _{I,\rm core}-\phi _{B}$, and $\phi _{I,\rm core}- \phi _{I,\rm cont}$ derived from Eq. (\ref{equa7}).

\begin{figure}
\epsscale{1.0}
\plotone{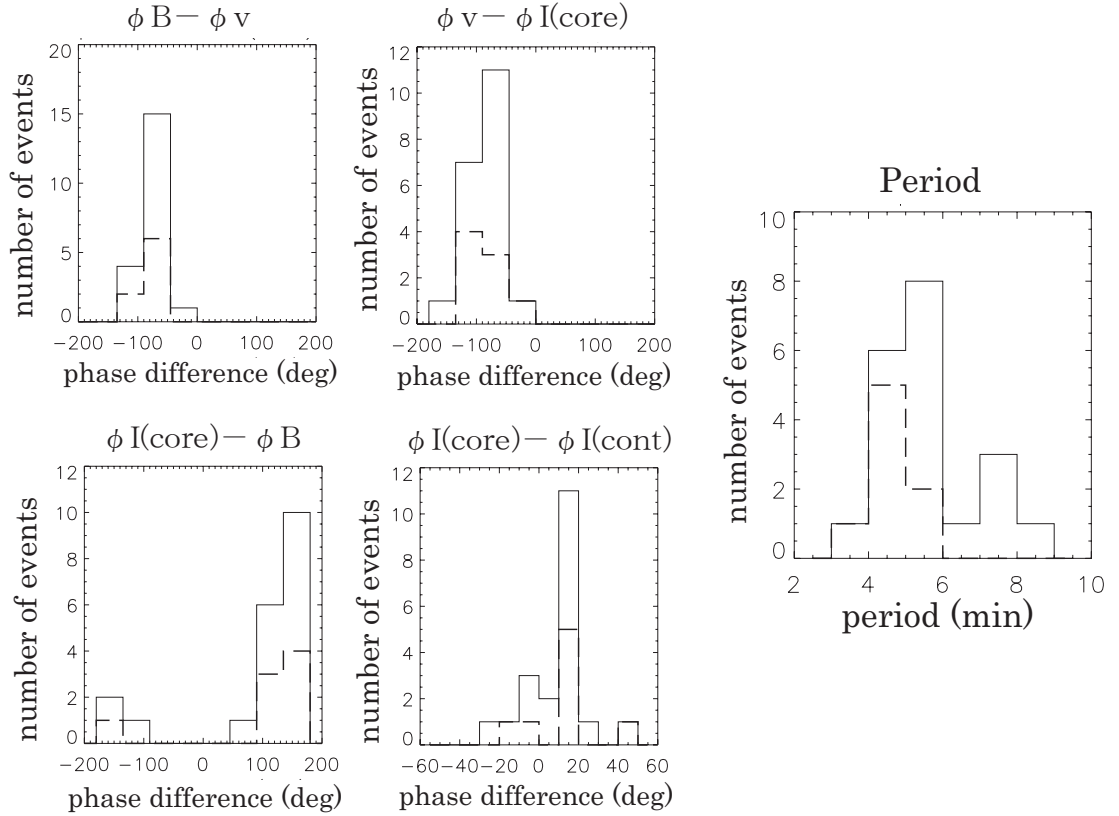}
\caption{
\emph{Left}: Histograms of the phase difference between fluctuations in the magnetic flux, the velocity, the line-core intensity, and the continuum intensity, $\phi _{B}-\phi _{v}$ $(\emph{top, left})$, $\phi _{v}-\phi _{I,\rm core}$ (\emph{top right}), $\phi _{I,\rm core}-\phi _{B}$ (\emph{bottom left}), and $\phi _{I,\rm core}-\phi _{I,\rm cont}$ (\emph{bottom right}). Solid lines indicate the phase difference for pores and IMSs, while dashed lines the phase difference for pores only. The histograms show striking concentrations at  around $-90^{\circ}$ for $\phi _{B}-\phi _{v}$ and $\phi _{v}-\phi _{I,\rm core}$, at around $180^{\circ}$ for $\phi _{I,\rm core}-\phi _{B}$, and at around $10^{\circ}$ for $\phi _{I,\rm core}-\phi _{I,\rm cont}$. \emph{Right}: Histogram of the periods of the common peaks in the power spectra. The peak periods are around 3$-$5 minutes for pores, while the peak periods for IMSs are around 4$-$9 minutes. 
}
\label{fig5}
\end{figure}

There are 8 cases for pores and 12 cases for IMSs where magnetic, velocity, and intensity fluctuations have strong power at the same periods. The histograms of the phase difference and period for 20 such common peaks are shown in Fig. \ref{fig5}. The histograms for the phase difference show striking concentrations at  around $-90^{\circ}$ for $\phi _{B}-\phi _{v}$ and $\phi _{v}-\phi _{I,core}$, at around $180^{\circ}$ for $\phi _{I,core}-\phi _{B}$, and at around $10^{\circ}$ for $\phi _{I,core}- \phi _{I,cont}$. Here, for instance, $\phi _{B}-\phi _{v} \sim -90^{\circ}$ means that the velocity leads the magnetic field by a quarter of cycle.
The periods are around 3$-$5 min for pores, while the periods are around 4$-$9 min for IMSs. There is no power between 134s (the detection limit due to the Nyquist criteria, see section 2.1.) and 204s (region \#04 in Table \ref{table2}). 

As pointed out in Sect. 2.2, no cross-talk should be expected in the l.o.s. magnetic signal from the velocity fluctuations. Furthermore, the phase difference between the magnetic flux and the velocity fluctuation $\phi _{B} - \phi _{v}$, if caused by the cross-talk, should be $0^{\circ}$ or $180^{\circ}$, while the observed phase difference shows a strong concentration at around $-90^{\circ}$. A similar phase relation is obtained by Bellot Rubio et al. (2000) for sunspot umbrae. On the other hand, R\"{u}edi et al. (1999) and Norton et al. (1999) came to an opposite conclusion that the magnetic field leads the velocity by about a quarter of a cycle.

\begin{landscape}
\begin{table} [htbp]
\tiny
\begin{center}
\tablenotetext{1}{r.m.s. (root mean square) l.o.s. (line of sight) magnetic flux amplitude}
\tablenotetext{2}{l.o.s. magnetic flux from Milne-Eddington inversion}
\tablenotetext{3}{average filling factor}
\tablenotetext{4}{r.m.s. l.o.s. velocity amplitude}
\tablenotetext{5}{r.m.s. intensity fluctuation normalized by the average intensity for line core and continuum}
\tablenotetext{6}{period}
\tablenotetext{7}{phase difference between magnetic, velocity, line core intensity, and continuum intensity fluctuations}
\begin{tabular}{ccccccccccccc} 
\hline
Region & $\delta \Phi_{\rm los,rms}$\tablenotemark{1} & $\Phi_{0,\rm los}$\tablenotemark{2} &$\frac{\delta \Phi_{\rm los,rms}}{\Phi_{0,\rm los}}$ & $f$ \tablenotemark{3} & $\delta v_{\rm los,rms}$\tablenotemark{4} & $\frac{\delta I_{\rm core,rms}}{I_{\rm cont}}$\tablenotemark{5} & $\frac{\delta I_{\rm rms,cont}}{I_{\rm cont}}$\tablenotemark{5} & $P$ \tablenotemark{6} & $\phi _{B}-\phi _{v}$\tablenotemark{7} & $\phi _{v}-\phi _{I,\rm core}$\tablenotemark{7} & $\phi _{I,\rm core}-\phi _{B}$\tablenotemark{7} & $\phi _{I,\rm core}-\phi _{I,\rm cont}$ \\  
ID & (G) & (10$^{3}$G) & (\%) & & (m/s) & (\%) & (\%) & (min) &(deg) & (deg) & (deg) & (deg) \\ \hline
\#01 & 17.1 & 1.16 & 1.48 & 0.77 & 70  & 0.58 & 0.38 & 4.0 & $-67$  & $-103$ & 170    & $-3$  \\ 
\#02 & 8.8  & 1.05 & 0.84 & 0.75 & 60  & 0.32 & 0.17 & 5.2 & $-57$  & $-74$  & 131    & 43    \\ 
 -   & 8.3  &   -  & 0.79 &  -   & 68  & 0.55 & 0.25 & 4.9 & $-58$  & $-123$ & $-179$ & 16    \\ 
 -   & 9.4  &   -  & 0.90 &  -   & 57  & 0.59 & 0.29 & 4.0 & $-54$  & $-110$ & 164    & 15    \\ 
\#03 & 8.9  & 0.78 & 1.14 & 0.65 & 86  & 0.47 & 0.36 & 5.2 & $-74$  & $-70$  & 145    & $-5$  \\ 
\#04 & 10.0 & 1.02 & 0.98 & 0.73 & 36  & 0.57 & 0.28 & 3.4 & $-94$  & $-107$ & $-159$ & $-12$ \\ 
\#05 & 13.8 & 0.97 & 1.42 & 0.81 & 120 & 0.97 & 0.89 & 4.9 & $-57$  & $-73$  & 130    & 14    \\  
\#06 & 4.4  & 0.67 & 0.66 & 0.56 & 76  & 0.27 & 0.11 & 5.2 & $-71$  & $-47$  & 118    & 12    \\ 
\#07 & 9.8  & 1.08 & 0.91 & 0.72 & 67  & 0.36 & 0.34 & 7.6 & $-67$  & $-41$  & 108    & 19    \\ 
 -   & 7.7  &   -  & 0.71 &  -   & 59  & 0.35 & 0.25 & 4.3 & $-96$  & $-76$  & 172    & 11    \\ 
\#08 & 3.9  & 0.54 & 0.72 & 0.49 & 77  & 0.37 & 0.30 & 7.6 & $-58$  & $-61$  & 119    & 14    \\ 
 -   & 3.5  &   -  & 0.65 &  -   & 62  & 0.28 & 0.12 & 6.8 & $-58$  & $-120$ & 178    & 22    \\
\#09 & 4.8  & 0.46 & 1.04 & 0.46 & 98  & 0.74 & 0.53 & 5.7 & $-60$  & $-85$  & 145    & 14    \\ 
\#10 & 4.5  & 0.51 & 0.88 & 0.57 & 34  & 0.19 & 0.12 & 7.6 & $-105$ & $-156$ & $-99$  & $-21$ \\
\#11 & 7.4  & 0.47 & 1.57 & 0.52 & 35  & 0.92 & 0.58 & 7.6 & $-102$ & $-87$  & 179    & 16    \\ 
\#12 & 4.5  & 0.58 & 0.78 & 0.53 & 44  & 0.25 & 0.15 & 5.7 & $-38$  & $-46$  & 84     & 16    \\
 -   & 3.5  &   -  & 0.60 &   -  & 82  & 0.41 & 0.30 & 5.2 & $-48$  & $-100$ & 148    & 8     \\
\#13 & 5.1  & 0.73 & 0.70 & 0.61 & 73  & 0.39 & 0.18 & 5.2 & $-48$  & $-71$  & 118    & 4     \\
 -   & 6.4  &   -  & 0.88 &  -   & 47  & 0.47 & 0.26 & 4.5 & $-55$  & $-93$  & 138    & 21    \\     
\#14 & 6.4  & 0.39 & 1.64 & 0.43 & 40  & 0.20 & 0.11 & 8.5 & $-77$  & $-88$  & 165    & -7    \\ 
\hline
\end{tabular}
\caption{
Physical parameters corresponding to the principal peak in the power spectra of all region of interests (shown in Table. \ref{table1}) with common peaks in the magnetic flux, the velocity, and the photometric intensity. 
} 
\label{table2}
\end{center}
\end{table}
\end{landscape}  

\section{Intensity Fluctuation}
Previous authors (e.g. Bellot Rubio et al. 2000) detected fluctuations in the magnetic field strength and the velocity for a sunspot umbra, and obtained a phase difference of $\sim 90^{\circ}$. They concluded that the observed fluctuations in magnetic field strength is mainly caused by the opacity effect. Temperature and density fluctuations associated with the propagation of a hydrodynamic (acoustic) or magneto-hydrodynamic (magneto-acoustic)  wave may cause the opacity fluctuation that moves the line formation layer upward or downward, resulting in an apparent magnetic field fluctuation, if the magnetic field has a gradient with geometrical height $(dB/dz)$. This is called the opacity effect. 

In this section, we consider whether the observed fluctuation is due to the opacity effect. The photometric intensity that we observe is given by
\begin{eqnarray}
I = \int_{0}^{\tau} \frac{\sigma T(\tau)^4}{\pi} e^{-\tau} d\tau \label{equa8}, 
\end{eqnarray} 
where $T$ is the local temperature at the optical depth $\tau$. The intensity modulation can take place due either to change in the temperature or to change in the optical depth, which depends on the density and the temperature in the optical path. The opacity effect involves the second term ($e^{-\tau}$). Fluctuation in intensity indicates a compressive nature of the fluctuation due to the first term ($\frac{\sigma T(\tau)^4}{\pi}$) and/or to the second term ($e^{-\tau}$) in eq. (\ref{equa8}).  Thus, waves with low intensity fluctuation, especially those with an intensity fluctuation close to zero, can be considered to be a incompressible mode (such as the kink mode), while those with high intensity fluctuation can be considered to be a compressible mode (such as the sausage mode).  

\begin{figure}
\epsscale{0.7}
\plotone{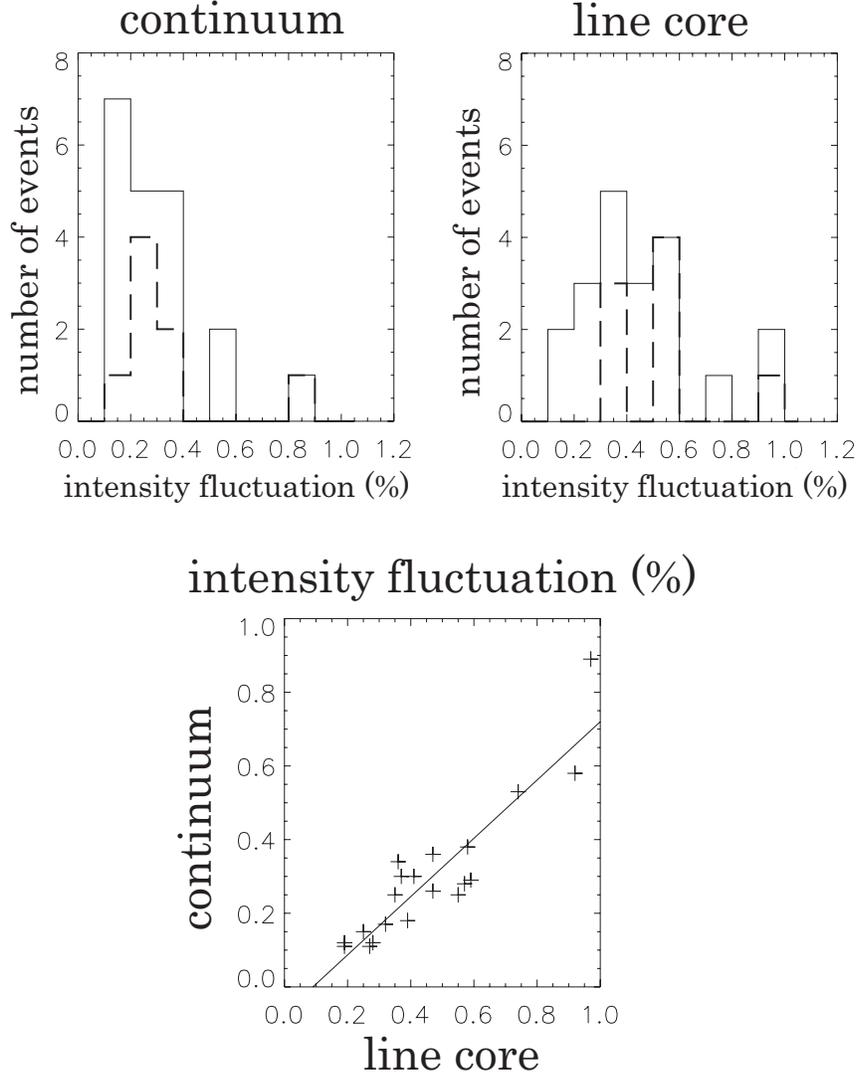}
\caption{
\emph{Top}: Histograms of continuum $\delta I_{\rm cont}$ and line core $\delta I_{\rm core}$ intensity fluctuations normalized by the average intensity $\overline{I_{\rm cont}}$ and $\overline{I_{\rm core}}$, $\frac{\delta I_{\rm cont,rms}}{\overline{I_{\rm cont}}}$ \emph{(left)} and $\frac{\delta I_{\rm core,rms}}{\overline{I_{\rm core}}}$ \emph{(right)} (solid lines). Dashed lines indicate histograms for pores. \emph{Bottom}: Scatter plot between the intensity fluctuations $\frac{\delta I_{\rm cont,rms}}{\overline{I_{\rm cont}}}$ and $\frac{\delta I_{\rm core,rms}}{\overline{I_{\rm core}}}$. The solid line indicates the linear regression line.
}
\label{fig6}
\end{figure}

The top panels of Fig. \ref{fig6} show the histograms of the line core $(\delta I_{\rm core,rms})$ and the continuum $(\delta I_{\rm cont,rms})$ intensity fluctuations normalized by the average intensity $\overline{I_{\rm core}}$ and $\overline{I_{\rm cont}}$; $\frac{\delta I_{\rm core,rms}}{\overline{I_{\rm core}}}$ (core fluctuation) and $\frac{\delta I_{\rm cont,rms}}{\overline{I_{\rm cont}}}$ (continuum fluctuation) for all the peaks. The relation between the core and the continuum fluctuations is shown in the bottom panel of Fig. \ref{fig6}. The scatter plot indicates that the fluctuation at the line core is larger than the continuum fluctuation for all the peaks, and that the line-core and the continuum fluctuations are linearly correlated. A linear fitting between the line-core and the continuum fluctuations is given by,
\begin{eqnarray}
\frac{\delta I_{\rm cont,rms}}{\overline{I_{\rm cont}}}=0.79\frac{\delta I_{\rm core,rms}}{\overline{I_{\rm core}}}-0.00066. \label{equa9}
\end{eqnarray}
The cross correlation coefficient is 0.91. Fig. \ref{fig5} shows that phase difference between the intensity fluctuation in the core and in the continuum, $(\phi _{I,\rm core}- \phi _{I,\rm cont})$, has a concentration at around $10^{\circ}\pm 14^{\circ}$.

We here consider the opacity effect due either to the density or to the temperature fluctuations. First we assume only the density fluctuation (without the temperature fluctuation). Magnetic field strength is smaller with height ($dB/dz<0$) because of the canopy structure of magnetic flux tubes. Since the observations are carried out with 25.2 to 49 degrees away from the normal, we simply  assume here that the magnetic field strength along the line of sight decreases with height in the following discussion. The temperature is lower with height below the temperature minimum. When the atmosphere in the line formation layer is compressed (or decompressed), the line formation layer moves upward (downward), because the opacity along the line of sight in the flux tube increases (decreases). When the line formation layer moves upward (or downward), both the magnetic field strength and the intensity decrease (increase). Therefore, the observed magnetic field strength and the constant-temperature intensity fluctuation caused by the opacity effect should have had the phase difference of $0^{\circ}$, while the observed phase differences $\phi _{I,core}- \phi _{B}$ have a concentration at around $180^{\circ}$. Thus, the observed phase difference is not  consistent with that caused by the opacity effect, if the opacity effect is caused only by the density fluctuation without temperature fluctuation.

On the other hand, the line formation layer may be compressed (or decompressed) under the adiabatic condition. We here consider the opacity effect due to temperature, assuming that the optical depth $\tau$ depends only on the temperature. The dominant absorber in the visible wavelengths is the H$^{-}$ ion (e.g. Stix, 2002). The populations of H$^{-}$ and HI are related with the Saha equation (Rutten 1995, eq. (8.2))
\begin{eqnarray}
\log \frac{N(\rm HI)}{N(\rm H^{-})}=-0.1761-\log P_{e}+\log \frac{U(\rm HI)}{U(\rm H^{-})}+2.5 \log T_{e}-\frac{5040\chi}{T_{e}}, \label{equa10}
\end{eqnarray}
where $P_{e}$ is the electron pressure, $T_{e}$ the electron temperature, $\chi$ the ionization energy from H$^{-}$ to H, $N(\rm H^{-})$ and $N(\rm HI)$ the population densities of H$^{-}$ and HI, $U(\rm H^{-})$ and $U(\rm HI)$ the partition function of H$^{-}$ and HI. 
Equation (\ref{equa10}) indicates that the population of H$^{-}$ depends highly on the temperature, and decreases with the temperature in the case of the adiabatic compression, while the population depends weakly on the pressure, and increases with the pressure in the constant temperature case. Thus, we cannot determine the population of H$^{-}$ in the actual situation without employing a model taken into account the radiation exchange between the inside and the outside of the flux tubes. 

We point out that regardless of mechanism to change the opacity, the phase difference between the fluctuations in the magnetic field and the intensity ($\phi_{I}-\phi_{B}$) depends only on the sign of magnetic gradient along the line of sight when the line formation height moves upward or downward due to the lateral expansion and the contraction of the tube. The flux tubes that we observed were located 25.2 to 49.0 degrees away from the sun center. If the magnetic field strength decreases with height along the oblique line of sight, the phase difference between the fluctuations in the magnetic field and the intensity $\phi_{I}-\phi_{B}$ should have been $0^{\circ}$, whereas we obtained $\phi_{I} - \phi_{B} \sim 180^{\circ}$. Therefore, the phase relation between the fluctuations in the magnetic field and the intensity from the observation would not be consistent with that caused by the opacity effect under the assumption of the decreasing field strength with height along the line of sight.  

If the effect of the adiabatic compression (or decompression; first term in equation (\ref{equa8})) is larger than the opacity effect due to the density and/or temperature fluctuation (second term in equation (\ref{equa8})), the phase difference between the magnetic field strength and the intensity fluctuation is $0^{\circ}$ for the case of the fast-mode wave, while that is $180^{\circ}$ for the case of the slow-mode wave. Thus, we can rule out the fast-mode wave, since the observed phase difference is close to $180^{\circ}$. 

\begin{figure}
\epsscale{1.0}
\plotone{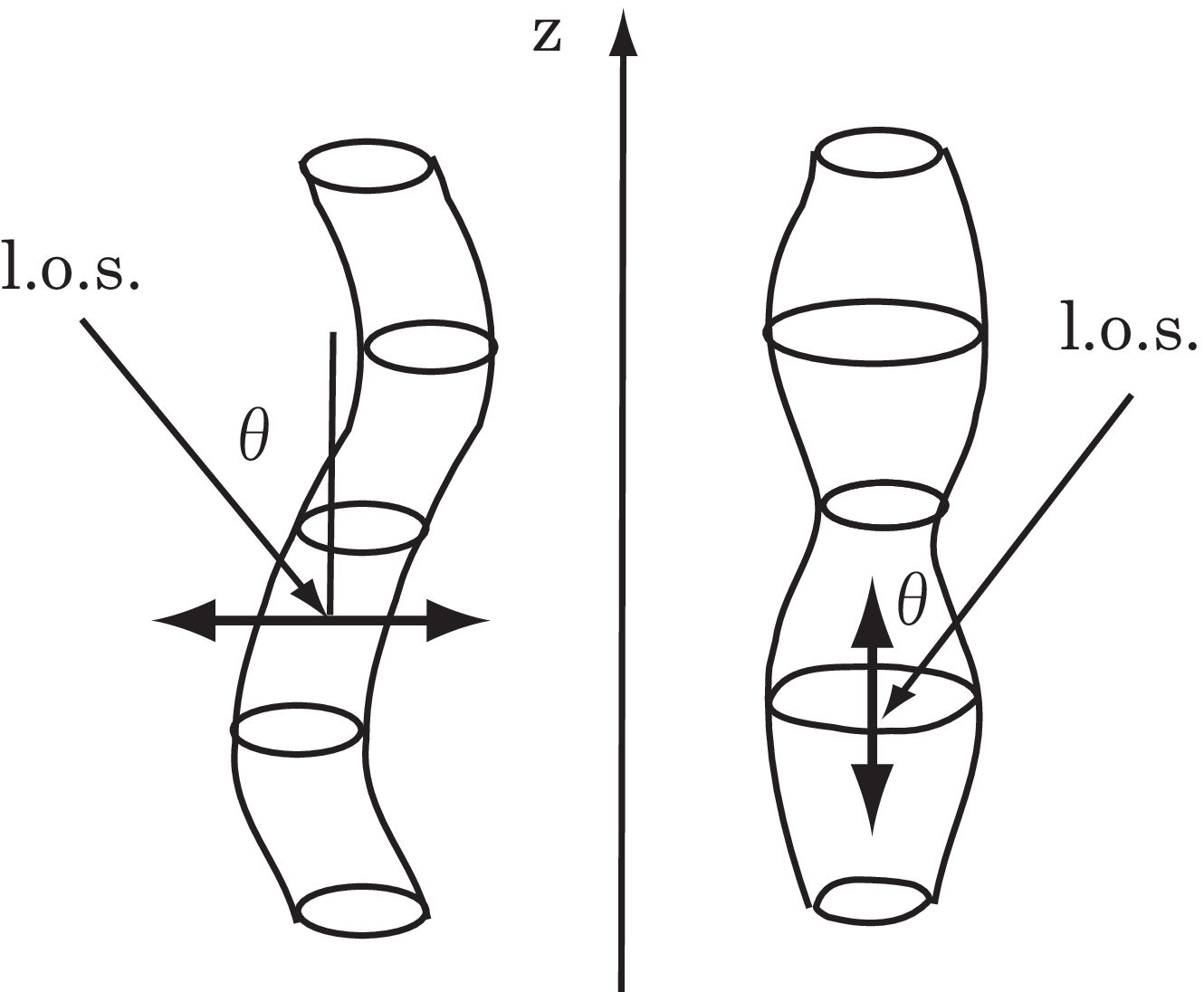}
\caption{
\emph{Left}:kink mode MHD wave \emph{Right}: sausage mode MHD wave
}
\label{fig7}

\end{figure}
\section{Kink Mode MHD Waves}
In this chapter, we examine whether the observed properties of waves are consistent with the kink mode MHD waves (Fig. \ref{fig7}). Though the magnetic and velocity fluctuations that we observe could be either parallel or perpendicular to the flux tubes, we here consider the possibility that the observed fluctuations are transverse to the magnetic field. As discussed in chapter 4, Fig. \ref{fig6} shows that some of the fluctuations has very small intensity fluctuation. Since the kink mode is essentially of non-compressive nature, those fluctuations with little intensity fluctuation may dominantly have properties of the kind mode. 

\subsection{Reflection of Kink waves}
The dispersion relation of the kink mode neglecting gravitational stratification is given by (e.g. Spruit, 1981; Edwin and Roberts, 1983, Moreno-Insertis, Sch\"{u}ssler, $\&$ Ferriz-Mas, 1996, Ryutova  \& Khijakadze, 1990)
\begin{eqnarray}
c_{k}=\frac{\omega}{k}=v_{A}\sqrt{\frac{\rho _{i}}{\rho _{i}+\rho _{e}}}, \label{equa11}
\end{eqnarray}
where $c_{k}$ is the phase speed of the kink mode, $\omega$ the frequency, $k$ the wave number, $v_{A}$ the Alfv\'en speed, $\rho _{i}$ the density inside the flux tube, and $\rho _{e}$ the density outside the flux tube. The transverse displacement of the flux tube $\delta x$ with geometrical height $z$ and time $t$ can be expressed as $\delta x(z,t)=x_{0} \cos (\omega t \pm kz)$, where $x_{0}$ is the amplitude of the transverse displacement. The transverse magnetic field and velocity component are given by, 
\begin{eqnarray}
\delta B= B_{0} \frac{\partial (\delta x)}{\partial z}= \mp B_{0}k \sin (\omega t \pm kz), \label{equa12} \\
\delta v= \frac{\partial (\delta x)}{\partial t}= - \omega \sin (\omega t \pm kz), \label{equa13}
\end{eqnarray}    
where $B_{0}$ is the vertical magnetic field strength. From Eqs. (\ref{equa11}) $-$ (\ref{equa13}), we obtain
\begin{eqnarray}
\frac{\delta B}{B_{0}}=\pm \frac{\delta v}{\omega /k}, \label{equa14} \\
\delta B= \pm \sqrt{4 \pi (\rho _{i}+\rho _{e})} \delta v. \label{equa15} 
\end{eqnarray}
The phase relation of the kink mode is the same as that of the Alfv\'en mode. Magnetic field is directed away from the Sun in our case. If the kink wave propagates to the direction same as that of magnetic field vector, minus sign should be taken, and vice versa. If a pure ascending or descending kink wave propagates toward the observer along magnetic field, phase difference between the magnetic field and the velocity fluctuations $(\phi _{B} - \phi _{v})$ should, therefore, have been 180$^{\circ}$ or 0$^{\circ}$, respectively.  Fig. \ref{fig5} shows that this is not the case. 

We then consider a superposition of the ascending kink wave and the descending waves, which is the reflected ascending wave at the photosphere-chromosphere boundary. When the ascending and the descending kink waves coexist in the line formation layer, the superposed wave form is determined by six variables $\delta B_{u}, \delta v_{u}, \phi_{u}, \delta B_{d}, \delta v_{d}, \phi _{d}$, which indicate the amplitude of the magnetic field fluctuation ($\delta B$), the amplitude of the velocity field fluctuation ($\delta v$), and the initial phase ($\phi$) of upward (subscript $u$) and downward (subscript $d$) waves. When magnetic field vector is toward the observer, the transverse magnetic field and velocity displacement of the superposed kink wave are given by
\begin{eqnarray}
\delta B=-\delta B_{u} \cos (\omega t+\phi _{u})+\delta B_{d} \cos (-(\omega t +\phi _{d})), \label{equa16} \\
\delta v= \delta v_{u} \cos (\omega t+\phi _{u})+\delta v_{d} \cos (-(\omega t +\phi _{d})). \label{equa17}
\end{eqnarray} 
Note that the phase difference between magnetic and velocity fluctuation in the ascending kink wave is 180$^{\circ}$, while that in the descending kink wave is 0$^{\circ}$. This fact is reflected in the sign of each term in Eqs. (\ref{equa16}) and (\ref{equa17}). We can rewrite these equations as follows:
\begin{eqnarray}
\delta B=\delta B_{s} \cos (\omega t+\phi _{B}), \label{equa18} \\
\delta v=\delta v_{s} \cos (\omega t+\phi _{v}), \label{equa19}
\end{eqnarray}
where $\delta B_{s}$ and $\delta v_{s}$ are the magnetic and the velocity amplitudes of the superposed kink wave, and $\phi _{B}$ and $\phi _{v}$ are phases of the magnetic field and the velocity of the superposed kink wave. In Eqs. (\ref{equa18}) and (\ref{equa19}), $\delta B_{s}$, $\delta v_{s}$, $\phi _{B}$, and $\phi _{v}$ are given by:
\begin{eqnarray}
\delta B_{s}=\sqrt{\delta B_{u}^{2}+\delta B_{d}^{2}-2\delta B_{u} \delta B_{v}\cos (\phi _{u}-\phi _{d})}, \label {equa20} \\
\cos \phi _{B}=\frac{\delta B_{u}\sin \phi _{u}-\delta B_{d}\sin \phi _{d}}{\delta B_{s}}, \label {equa21} \\
\sin \phi _{B}=\frac{-\delta B_{u}\cos \phi _{u}+\delta B_{d}\cos \phi _{d}}{\delta B_{s}}, \label {equa22} \\
\delta v_{s}=\sqrt{\delta v_{u}^{2}+\delta v_{d}^{2}+2\delta v_{u}\delta v_{v}\cos (\phi _{u}-\phi _{d})}, \label {equa23} \\
\cos \phi _{v}=\frac{-\delta v_{u}\sin \phi _{u}-\delta v_{d}\sin \phi _{d}}{\delta v_{s}}, \label {equa24} \\
\sin \phi _{v}=\frac{\delta v_{u}\cos \phi _{u}+\delta v_{d}\cos \phi _{d}}{\delta v_{s}}. \label {equa25}
\end{eqnarray}
>From Eq. (\ref{equa15}), we obtain $\frac{\delta v}{\delta B}=\frac{1}{\sqrt{4\pi (\rho _{i}+ \rho _{e})}}$. Therefore, we obtain the following relation among the quantities in Eqs. (\ref{equa16}) and (\ref{equa17}): 
\begin{eqnarray}
\frac{\delta v_{u}}{\delta B_{u}}=\frac{\delta v_{d}}{\delta B_{d}}. \label{equa26} 
\end{eqnarray} 
Using Eqs. (\ref{equa20})$-$(\ref{equa26}), the following phase difference between magnetic and velocity fluctuations is obtained:
\begin{eqnarray}
\cos (\phi_{B}-\phi_{v})=\cos \phi _{B} \cos \phi _{v}+ \sin \phi _{B} \sin \phi _{v} \nonumber \\
=\frac{-\delta B_{u}\delta v_{u}+\delta B_{d}\delta v_{d}}{\delta B_{s}\delta v_{s}} = \frac{\delta B_{u}/\delta v_{u}}{\delta B_{s}\delta v_{s}} (-\delta v_{u}^{2}+\delta v_{d}^{2})= \frac{\delta v_{u}/ \delta B_{u}}{\delta B_{s} \delta v_{s}} (-\delta B_{u}^{2}+\delta B_{d}^{2}). \label{equa27}
\end{eqnarray}
This equation shows that the phase difference between the magnetic and the velocity fluctuations ($\phi _{B}- \phi _{v}$) should be $-90^{\circ}$ or $90^{\circ}$ when the amplitude of the reflected descending kink wave is exactly the same as that of ascending kink wave (i.e. $\delta v_{u}=\delta v_{d}$ and $\delta B_{u}=\delta B_{d}$). The observed phase relation is consistent with this prediction.

\subsection{Standing kink waves}
The transverse displacement of magnetic field line in the presence of upward ($\delta x_{u}$) and downward ($\delta x_{d}$) kink wave is written as a function of height $(z)$ and time $(t)$,
\begin{eqnarray}
\delta x_{u}(t,z)=x_{u0} \cos (\omega t+kz+\phi _{u}) \label{equa28}, \\
\delta x_{d}(t,z)=x_{d0} \cos (\omega t-kz+\phi _{d}) \label{equa29},
\end{eqnarray}
where $x_{u0}, x_{d0}, \phi _{u}, \phi _{d}$ are the transverse amplitude and the initial phase of the magnetic field line fluctuation in the presence of the upward (subscript $u$) and the downward (subscript $d$) kink wave. When $x_{u0}=x_{d0} \equiv x_{0}$, which corresponds to the case for perfect reflection, the transverse displacement $\delta x_{s}$ of the magnetic field line in the presence of the superposed kink waves is given by
\begin{eqnarray}
\delta x_{s}(t,z)=\delta x_{u}(t,z)+\delta x_{d}(t,z)=2x_{0} \cos (\omega t+\frac{\phi _{u}+\phi _{d}}{2}) \cos (kz+\frac{\phi _{u}-\phi _{d}}{2}). \label{equa30}
\end{eqnarray}
Equation (\ref{equa30}) shows that the superposed kink wave, if with perfect reflector, is a standing wave. Sketches of standing kink wave are shown in Fig. \ref{fig8}. Whether the phase difference is  $90^{\circ}$ or $-90^{\circ}$ depends on the distance from the reflection boundary (node).

\subsection{Phase Difference}
\begin{figure}
\epsscale{0.6}
\plotone{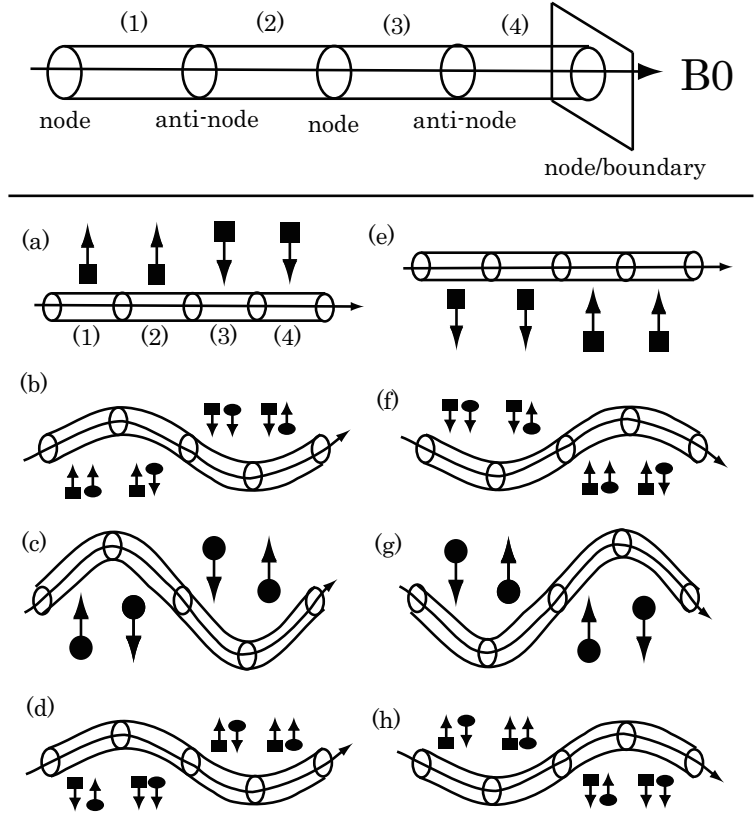}
\caption{
\emph{Top}: A standing kink wave along magnetic field line $B0$ is divided into four parts (1) through (4) each separated by nodes and anti-nodes. \emph{Bottom}: Time evolution of the standing kink waves. The wave evolves from (a) to (h), and goes back to (a). The arrows with filled box indicate velocity vector, while the arrows with circle indicate perturbed component of magnetic field vector. The length of the arrows indicates the magnitudes of the vector at certain space and time points. Schematic representation of the standing kink wave shows that the phase difference between magnetic and velocity fluctuations ($\phi _{B}-\phi _{v}$) is $-90^{\circ}$ at the portions (1) and (3), and $90^{\circ}$ at the portions (2) and (4).    
}
\label{fig8}
\end{figure}

We here give one interpretation for the concentration of the phase difference at around $-90^{\circ}$ (Fig. \ref{fig8}). When the ascending kink wave is reflected back at chromosphere-corona boundary, and the ascending and the descending kink waves coexist in the line formation layer beneath the reflector, the phase difference between the magnetic and the velocity fluctuations should have been either $90^{\circ}$ or $-90^{\circ}$, while observed phase angle concentrates at around  $-90^{\circ}$. Whether the phase angle is $90^{\circ}$ or $-90^{\circ}$ depends on the distance between the reflector and the line formation layer (Fig. \ref{fig8}). The concentration at $-90^{\circ}$ indicates that the separation between the reflecting boundary and the line formation layer is fixed for all the flux tubes such that it corresponds to $-90^{\circ}$ phase difference. If we perform similar observations with different absorption lines with different formation height, and the difference in height is larger than the quarter of the wavelength (800km), this conjecture can be verified.

\subsection{Leakage of Poynting Flux to Corona}
The Poynting flux above the reflecting layer is the Poynting flux of the ascending kink wave minus the Poynting flux of the descending kink wave in the line formation layer. We here estimate the effective or residual upward-directed Poynting flux along a flux tube above the reflector. The Poynting flux of the kink wave is given by $F=\frac{\overline{f} B_{0}}{4\pi}(\delta B_{\rm rms} \delta v_{\rm rms})$, so that the difference of the Poynting flux between the ascending and the descending kink waves is given by
\begin{eqnarray}
\triangle F=\frac{\overline{f} B_{0}}{4\pi}(\delta B_{u,\rm rms}\delta v_{u,\rm rms}-\delta B_{d,\rm rms}\delta v_{d,\rm rms}), \label{equa31}
\end{eqnarray}
where $\delta B_{u,\rm rms}=\delta B_{u}/\sqrt{2}$, $\delta v_{u,\rm rms}=\delta v_{u}/\sqrt{2}$, $\delta B_{d,\rm rms}=\delta B_{d}/\sqrt{2}$, $\delta v_{d,\rm rms}=\delta v_{d}/\sqrt{2}$. Using Eq. (\ref{equa27}), we can rewrite the equation as follows:
\begin{eqnarray}
\triangle F=-\frac{\overline{f} B_{0}}{4\pi}(\delta B_{s,\rm rms} \delta v_{s,\rm rms})\cos (\phi _{B}-\phi _{v}), \label{equa32}
\end{eqnarray}
where $\delta B_{s,\rm rms}=\delta B_{s}/\sqrt{2}$, $\delta v_{s,\rm rms}=\delta v_{s}/\sqrt{2}$. It turns out that the effective upward-directed Poynting flux is proportional to $\cos (\phi _{B}-\phi _{v})$. $\delta B_{s,\rm rms}$, and $\delta v_{s,\rm rms}$ in Eq. (\ref{equa32}) are related to the observables, assuming that the fluctuations are transverse (i.e. normal to the flux tubes), 
\begin{eqnarray}
\delta B_{s,\rm rms}=\frac{\delta \Phi_{\rm los,rms}}{\overline{f} \sin \theta}, \label{equa33} \\
\delta v_{s,\rm rms}=\frac{\delta v_{\rm los,rms}}{\sin \theta}, \label{equa34} 
\end{eqnarray}
where $\theta$ is helio-longitudinal angle from the meridional line. If the phase difference from $-90^{\circ}$ is just 6$^{\circ}$ as an exercise, i.e. $\phi _{B}-\phi _{v}=-96^{\circ}$, we obtain $\triangle F= 2.7 \times 10^{6}$ erg cm$^{-2}$ s$^{-1}$ by substituting $B_{0}=1.7 \times 10^{3}$ G, $\delta \Phi_{\rm los,rms}=7.7$ G, $\delta v_{\rm los,rms}=0.059$ km/s, $\overline{f}=0.73$, and $\theta = 29^{\circ}$ (region \#07). Therefore, even if we observe the considerable reflected wave in the photospheric layer with SOT/SP, there will be substantial leakage of the upward kink wave toward chromosphere and corona in terms of the energy flux required for the coronal heating ($\sim 3 \times 10^{5}$ erg cm$^{-2}$ s$^{-1}$ for the quiet Sun; Withbroe \& Noyes, 1977).  

\subsection{Seismology of Photospheric Flux Tubes}
We show in this chapter that various physical parameters that characterize the magnetic flux tubes are obtained simply from the amplitude and period of the magnetic and velocity fluctuations. We estimate the physical parameters for the region \#02. The intensity fluctuation is 0.17$-$0.25\% in continuum (Table \ref{table2}), and we assume that the observed fluctuation is due to the superposition of upward and downward kink waves. 

\begin{figure}
\epsscale{0.8}
\plotone{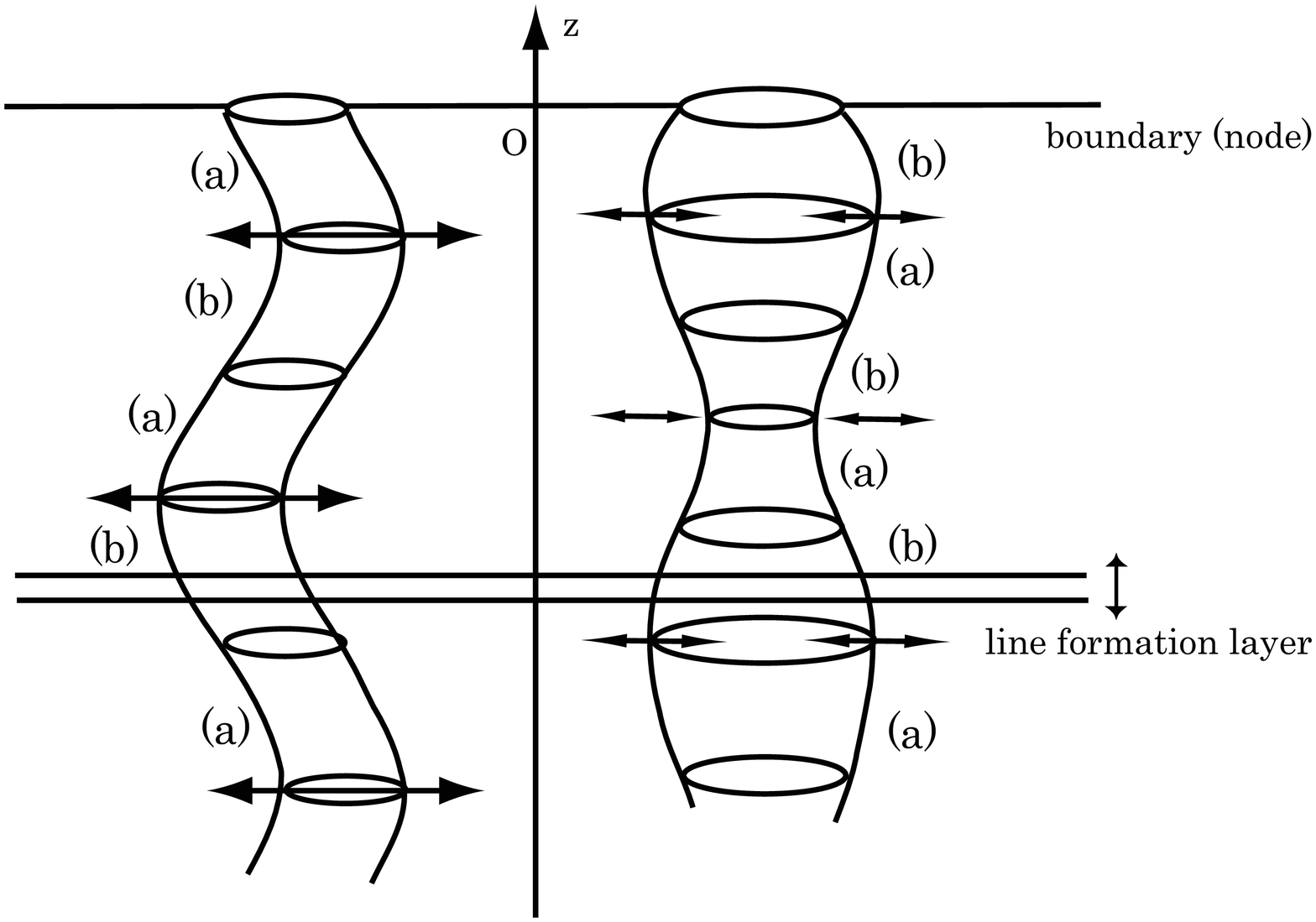}
\caption{
The standing kink wave (\emph{left}) and the standing slow sausage wave (\emph{right}). The phase difference between the fluctuations in the magnetic field and the velocity ($\phi_{B} - \phi_{v}$) is 90$^{\circ}$ in the sector (a) and $-$90$^{\circ}$ in the sector (b). The arrows indicate the transverse motion of the magnetic fields at the anti-nodes.
}
\label{fig9}
\end{figure}

We define the coronal/chromospheric boundary, which is considered to be a reflector, to be the origin of the z-axis, which is normal to the solar surface (away from the Sun). A schematic behavior of the standing kink wave is shown in the left panel of Fig. \ref{fig9}. Substituting $\frac{\phi_{u}+\phi_{d}}{2}=0$ (without losing generality) and $\frac{\phi_{u}-\phi_{d}}{2}=\frac{\pi}{2}$ (to make the height at $z=0$ the node) into Eq. (\ref{equa30}), the transverse displacement of the flux tube is given by,
\begin{eqnarray}
\delta x_{s}(t,z)=2x_{0}\cos (\omega t)\sin (kz). \label{equa35}
\end{eqnarray}
The transverse components of the magnetic field and the velocity are given by,
\begin{eqnarray}
\delta B_{s}(t,z)=B_{0} \frac{\partial \delta x_{s}}{\partial z}=2 B_{0} x_{0} k \cos (\omega t) \cos (kz), \label{equa36} \\
\delta v_{s}(t,z)=\frac{\partial \delta x_{s}}{\partial t}=-2 x_{0} \omega \sin (\omega t) \sin (kz), \label{equa37}
\end{eqnarray}
Equations (\ref{equa36}) and (\ref{equa37}) indicate that the phase difference between the fluctuations in the magnetic field and the velocity $\phi_{B} - \phi_{v}$ is,
\begin{eqnarray}
\left\{
\begin{array}{l}
90^{\circ} \ {\rm for} \ (n+\frac{1}{2}) \pi  \leq kz \leq (n+1) \pi  \ ({\rm sector \ (a) \ in \ Fig. \ref{fig9}}), \label{equa38} \\
-90^{\circ} \ {\rm for} \ n \pi  \leq kz \leq (n+\frac{1}{2}) \pi  \ ({\rm sector \ (b) \ in \ Fig. \ref{fig9}}), 
\end{array}
\right.
\end{eqnarray}
where {$n = -1, -2, -3, ...$}. Equation (\ref{equa38}) indicates that the observed phase difference $\phi_{B} - \phi_{v} \sim -90^{\circ}$ is consistent with the situation that the line-formation height is located in the sector (b). From Eqs. (\ref{equa11}), (\ref{equa36}), (\ref{equa37}), we have
\begin{eqnarray}
\frac{|\delta v_{s}|}{|\delta B_{s}|}=\frac{\omega /k}{B_{0}}|\tan (kz)|=\frac{|\tan (kz)|}{\sqrt{4 \pi (\rho _{i}+\rho_{e})}}, \label{equa39} \\
\rho_{i}+\rho_{e}= \Bigl( \frac{|\delta B_{s}|}{|\delta v_{s}|} \Bigr) ^{2} \frac{|\tan (kz)|^{2}}{4 \pi}, \label{equa40}
\end{eqnarray}
where $|\delta B_{s}|$ and $|\delta v_{s}|$ are the amplitude of the fluctuations in the magnetic field and the velocity, and are the function of height $z$. $|\delta B_{s}|$ and $|\delta v_{s}|$ in Eq. (\ref{equa40}) are related to the observables,
\begin{eqnarray}
|\delta B_{s}|=\frac{\sqrt{2}\delta \Phi_{\rm los,rms}}{\overline{f} \sin \theta}, \label{equa41} \\
|\delta v_{s}|=\frac{\sqrt{2}\delta v_{\rm los,rms}}{\sin \theta}. \label{equa42}   
\end{eqnarray}
Assuming that the flux tubes that we observe here are in pressure equilibrium, and do not have a helical structure (azimuthal component), the equation for the pressure equilibrium for the flux tube is simply expressed as 
\begin{eqnarray}
\frac{B^{2}_{i}}{8 \pi}+\frac{\rho _{i}}{m}k_{B} T_{i}=\frac{B^{2}_{e}}{8 \pi}+\frac{\rho _{e}}{m}k_{B} T_{e}, \label{equa43}\\
\rho _{e}T_{e}-\rho _{i}T_{i}=\frac{m}{8\pi k_{B}}(B_{i}^{2}-B_{e}^{2}), \label{equa44} 
\end{eqnarray}
where $B$, $\rho$, and $T$ are the magnetic field strength, the mass density and the temperature, and the  subscript \emph{i} and \emph{e} indicate the inside and the outside of the flux tube, respectively, $m$ average particle mass, and $k_{B}$ the Boltzmann constant. From Eqs. (\ref{equa40}) and (\ref{equa44}), we can determine $\rho_{i}$ and $\rho_{e}$ assuming that outside the flux tube is field-free ($B_{e}=0$ G), as is inferred by the observations.

The line formation height in the umbra is deeper than that in the quiet Sun, because of the lower temperature and density (e.g. Stix, 2002). The Wilson depression for the flux tube with $B \sim 2000$ G reaches about 300 $-$ 400 km (Deinzer, 1965; Mathew et al., 2004). The temperature and the average molecular weight at the height $\sim -350$ km is $T_{e}=1.0 \times 10^{4}$ K and $\mu = 1.2$ (from Table 2.4, Stix, 2002). Since the temperature inside the flux tube is lower than that outside the flux tube (Maltby et al, 1986), we assume $T_{i} = 7.0 \times 10^{3}$ K. We choose $kz=-496^{\circ}$ (see section 6.3 for justification to choose the value). Substituting $m=\mu m_{p}=1.9 \times 10^{-24}$ g, where $m_{p}$ is the proton mass, $k_{B}=1.4 \times 10^{-16}$ erg K$^{-1}$, $B_{i}=B_{0}=1.9 \times 10^{3}$ G, $\delta \Phi_{\rm los,rms}=8.8$ G, $\delta v_{\rm los,rms}=0.060$ km $s^{-1}$, $\overline{f}=0.75$, and $\theta=29^{\circ}$, we obtain mass densities $\rho _{i}=5.1 \times 10^{-8}$ g cm$^{-3}$ and $\rho _{e}=2.3 \times 10^{-7}$ g cm$^{-3}$. The number densities inside and outside the flux tube are $n_{i}=\frac{\rho_{i}}{m}=2.7 \times 10^{16}$ cm$^{-3}$ and $n_{e}=\frac{\rho_{e}}{m}=1.2 \times 10^{17}$ cm$^{-3}$, respectively. The mass density for the height of $-300 - -400$ km is $\rho_{e}=3.5 - 4.5 \times 10^{-7}$ g cm$^{-3}$ (from Table 6.1, Stix, 2002). This is consistent with our estimation within a factor of 2.

We also estimate other physical parameters associated with the flux tube: (1) Alfv\'en speed inside the flux tube $v_{A,i}=\frac{B_{i}}{\sqrt{4 \pi \rho_{i}}}$, (2) plasma $\beta =\frac{\rho_{i} k_{B}T/m}{B_{i}^{2}/8 \pi}$ inside the flux tube in the line formation layer, (3) wavelength of the kink mode $L=v_{A,i} \sqrt{\frac{\rho_{i}}{\rho_{i}+\rho_{e}}} P$, where $P$ is the fluctuation period, (4) propagation time of fast magneto-acoustic wave across the flux tube $\tau =\frac{R}{\sqrt{v_{A,i}^{2}+c_{s}^{2}}}$, where $R$ is the tube radius, and (5) distance between the boundary and the line formation layer $d=L\frac{|kz|}{360}$. Other obvious useful parameters are the pressure scale height $H=\frac{k_{B}T}{mg}$, where $g$ is the gravity in the solar surface, and the sound speed in the photosphere $c_{s}=\sqrt{\frac{\gamma k_{B}T}{m}}$, where $\gamma$ is the adiabatic coefficient. Substituting $B_{i}=1.7 \times 10^{3}$ G, $\rho _{i}=5.1 \times 10^{-8}$ g cm$^{-3}$, $\rho _{e}=2.3 \times 10^{-7}$ g cm$^{-3}$, $g=2.7 \times 10^{4}$ cm s$^{-2}$, $P=312$ s, $\gamma =5/3$, and $R=2000$ km (case \# 02), we obtain $v_{A,i}=24$ km s$^{-1}$, $\beta =0.18$, $L=3.1 \times 10^{3}$ km, $\tau =75$ s, $d=4.3 \times 10^{3}$ km, $H=3.9 \times 10^{2}$ km, and  $c_{s}=11$ km s$^{-1}$. The propagation time of the fast magneto-acoustic wave across the flux tube $\tau$ is less than the oscillation period $P$, and this is consistent with the assumption of the kink wave.

Mathew et al. (2004) calculated the physical parameters (magnetic pressure, gas pressure, Wilson depression, and plasma $\beta$) for a sunspot by performing an inversion to infrared spectro-polarimetric profiles, and derived plasma beta for the umbra $\beta \sim 0.5$. R\"{u}edi (1992) also performed an inversion to the infrared lines, and obtained the plasma $\beta \sim 0.25$ at $z=0$ km in the plage region. The plasma beta is generally higher at $z=-350$ km,  following the increase in the mass density (Stix, 2002).

As demonstrated here, we are potentially able to obtain all the physical parameters of the flux tube from the information on the MHD fluctuations. This indicates that seismology of magnetic flux tubes is possible with multiple lines corresponding to different height (photosphere and chromosphere) of the solar atmosphere.

\section{Sausage Mode MHD Waves}
We here consider the alternative possibility that the observed magnetic and velocity fluctuations are due to the longitudinal MHD waves or the slow sausage mode oscillation (Fig. \ref{fig7}: Ryutova, 2009; Defouw, 1976; Roberts and Webb, 1978; Ryutova 1981). 

\subsection{Phase Relation for Propagating Wave}
We consider a slow mode perturbation propagating along a cylindrical flux tube, neglecting gravitational stratification, following Ryutova (2009). We assume that the magnetic and velocity fluctuations with higher intensity fluctuation (Fig. \ref{fig6}) may have the sausage-mode nature. The momentum equation perpendicular to the flux tube is given by
\begin{eqnarray}
\frac{B_{0\|}\delta B_{\|}}{4\pi}+\delta p=0, \label{equa44.5}
\end{eqnarray}
where the subscript $0$ means these values in unperturbed state, and $\delta$ means perturbation of these values. We have the relation under the adiabatic condition
\begin{eqnarray}
\delta p = c_{s0}^2 \delta \rho , \label{equa45}
\end{eqnarray}
and the flux conservation is given by
\begin{eqnarray}
B_{0\|}\delta S +\delta B_{\|} S_0 =0. \label{equa46}
\end{eqnarray}
The momentum equation parallel to the flux tube is (substituting eq. (\ref{equa45}))
\begin{eqnarray}
\rho_0\frac{\partial \delta v_{\|}}{\partial t}= - \frac{\partial \delta p}
{\partial z}=  - c_{s0}^2\frac{\partial \delta \rho}{\partial z}, \label{equa47}
\end{eqnarray}
and the continuity equation is
\begin{eqnarray}
\frac{\partial}{\partial t}(\delta \rho S_0 + \delta S \rho_0) + S_0\rho_0
\frac{\partial \delta v_{\|}}{\partial z}=0, \label{equa48} 
\end{eqnarray} 
where $S=\pi R^2$, $B_{\|}$, $\rho$, $p$, $v_{\|}$ are the cross section of the flux tube, the longitudinal magnetic field, the density, the pressure, and the longitudinal velocity, respectively, and $c_{s,0}$ is the sound speed. From Eqs. (\ref{equa44}) $-$ (\ref{equa46}), we have
\begin{eqnarray}
\delta B_{\|}= -\frac{4\pi \delta p}{B_{0\|}}=-\frac{4\pi c_{s0}^2}{B_{0\|}}\delta \rho , \label{equa49} \\
\delta S = -S_0\frac{\delta B_{\|}}{B_{0\|}}= S_0\frac{4\pi c_{s0}^2}{B_{0\|}^2} \delta \rho . \label{equa50}
\end{eqnarray}
The continuity equation (Eq. \ref{equa48}) becomes
\begin{eqnarray}
\left (1+\frac{4\pi c_{s0}^2 \rho_0}{B_{0\|}^2}\right ) \frac{\partial \delta \rho}{\partial t} +\rho_0
\frac{\partial \delta v_{\|}}{\partial z}=0. \label{equa51}
\end{eqnarray}
Taking the time derivative, and substituting Eq. (\ref{equa47}), we have the dispersion relation, 
\begin{eqnarray}
\left (1+\frac{4\pi c_{s0}^2 \rho_0}{B_{0\|}^2}\right ) \frac{\partial^2 \delta \rho}{\partial t^2}-c_{s0}^2\frac{\partial^2 \delta \rho}{\partial z^2}=0. \label{equa52}  
\end{eqnarray}
We therefore obtain the phase velocity of the slow sausage mode $c_{T}$ (c.f. Edwin and Roberts, 1983), 
\begin{eqnarray}
c_{T}^{2}=\frac{c_{s0}^2 v_A^2}{c_{s0}^2 +v_A^2}, \label{equa53}
\end{eqnarray}
where $v_{A}$ is the Alfv\'en velocity. 

Hereafter we define positive as away from the solar surface. We consider a simple sinusoidal wave propagating upward ($k>0$) or downward ($k<0$) along the flux tube of positive ($B_{\|}>0$) or negative ($B_{\|}<0$) polarity,  
\begin{eqnarray}
\delta \rho=\delta \tilde{\rho}  ~\mbox{cos} (\omega t - k z) \qquad (\omega =kc_T), \label{equa54}
\end{eqnarray}
where $\delta \tilde{\rho}$ is the amplitude of the density fluctuation. Substituting Eq. (\ref{equa49}), we have
\begin{eqnarray}
\delta  B_{\|}= - \frac {4\pi c_{s0}^2 \delta \rho}{B_{\|}}=-\frac {4\pi c_{s0}^2}{B_{\|}} \delta \tilde{\rho}  ~\mbox{cos} (\omega t - k z), \label{equa55}
\end{eqnarray}
and we have from Eq. (\ref{equa47}) 
\begin{eqnarray}
\rho_0\frac{\partial \delta v_{\|}}{\partial t}=-c_{s0}^2 k \delta \tilde{\rho} ~\mbox{sin} (\omega t - k z). \label{equa56}
\end{eqnarray} 
Taking the integration with time (neglecting integration constant), we have 
\begin{eqnarray}
\delta v_{\|}=\frac{c_{s0}^2}{c_T} \frac{\delta \tilde{\rho}}{\rho_0} ~\mbox{cos} (\omega t - k z), \label{equa58} 
\end{eqnarray}
Assuming that the flux tube has an axis-symmetric sausage oscillation, the transverse velocity averaged over the whole pixels within the flux tube should be canceled out. Thus, what we detect as a clear strong peak in the l.o.s. velocity must be longitudinal, if the fluctuation is due to the propagating slow sausage mode. 

>From Eqs. (\ref{equa54}), (\ref{equa55}), and (\ref{equa58}), we have the phase relations between the fluctuations in the magnetic field, the velocity, and the density,
\begin{eqnarray}
\frac{\delta \rho}{\delta B_{\|}}=-\frac{B_{\|}}{4 \pi c_{s,0}^{2}}, \label{equa59} \\
\frac{\delta B_{\|}}{\delta v_{\|}}=-\frac{4 \pi c_{T}\rho_{0}}{B_{\|}}, \label{equa60} \\
\frac{\delta v_{\|}}{\delta \rho}=\frac{c_{s,0}^{2}}{c_{T}\rho_{0}}, \label{equa61} 
\end{eqnarray}
Equation (\ref{equa60}) indicates that the phase difference between the fluctuations in magnetic field and the velocity $\phi _{B}-\phi _{v}$ in the propagating wave with slow sausage mode is $0^{\circ}$ or $180^{\circ}$, depending on the direction of magnetic field and wave propagation, whereas we observed $\phi _{B} - \phi _{v} \sim -90^{\circ}$. Thus we can rule out the possibility that the observed fluctuations are due to the propagating wave with slow sausage mode.

\subsection{Phase Relation for the Standing Sausage Wave}
We here consider the superposition of ascending and the descending slow sausage waves with the same amplitude of the density fluctuation, assuming $B_{\|}>0$ from our observation, 
\begin{eqnarray}
\delta \rho= \delta \tilde{\rho} [\cos (k c_T t-kz+\phi_{u})+\cos (k c_T t+kz+ \phi_{d})] = \nonumber \\
\delta \tilde{\rho} ~\mbox{cos} (k c_T t+\frac{\phi_{u}+\phi_{d}}{2})  ~\mbox{cos}(k z+\frac{\phi_{u}-\phi_{d}}{2}), \label{equa62}
\end{eqnarray}
where $\phi_{u}$ and $\phi_{d}$ are the initial phases of the upward and downward propagating waves with slow sausage mode, and $k>0$ without losing generality. From Eqs. (\ref{equa49}) and (\ref{equa62}) we have
\begin{eqnarray}
\delta  B_{\|}= -\frac {4\pi c_{s0}^2}{B_{\|}} \delta \tilde{\rho} ~\mbox{cos} (k c_T t+\frac{\phi_{u}+\phi_{d}}{2})  ~\mbox{cos}(k z+\frac{\phi_{u}-\phi_{d}}{2}), \label{equa63}
\end{eqnarray}
and from Eqs. (\ref{equa47}) and (\ref{equa62}) we have
\begin{eqnarray}
\delta  v_{\|}=\frac { c_{s0}^2}{\rho c_T} \delta \tilde{\rho} ~\mbox{sin} (k c_T t+\frac{\phi_{u}+\phi_{d}}{2})  ~\mbox{sin}(k z+\frac{\phi_{u}-\phi_{d}}{2}). \label{equa64} 
\end{eqnarray} 
Equations (\ref{equa62}) and (\ref{equa63}) indicate that the phase difference between the fluctuations in the magnetic field and the density is $180^{\circ}$. 

Equations (\ref{equa63}) and (\ref{equa64}) indicate that the phase difference between the fluctuations in the magnetic field and the velocity $\phi_{B}-\phi_{v}$ is  $90^{\circ}$ or  $-90^{\circ}$, depending on the location of the line formation layer (right panel of Fig. \ref{fig9}). The observed phase relation between the fluctuations in the magnetic field and the velocity is  $-90^{\circ}$.

In the previous section, we discussed that waves with low intensity fluctuation be considered to be a incompressible mode (such as the kink mode), while those with high intensity fluctuation is considered to be a compressible mode (such as the sausage mode). However, Eq. (\ref{equa62}) indicates that the density fluctuation and the resultant intensity fluctuation are zero at the nodal points for the standing sausage wave. Thus, there may be cases that the standing sausage wave may not show intensity fluctuation with large amplitude. 

\subsection{Seismology of Photospheric Flux Tubes}
We here show that the seismology of magnetic flux tubes is also possible for the sausage MHD oscillation. We assume that the observed fluctuation is due to the superposition of upward and downward compressible sausage waves for the region \#05. This is justified by the fact that the region \#05 has very high intensity fluctuation (Table \ref{table2}). 

A schematic behavior of the standing sausage wave is shown in the right panel of Fig. \ref{fig9}. Substituting $\frac{\phi_{u}+\phi_{d}}{2}=0$ (without losing generality) and $\frac{\phi_{u}-\phi_{d}}{2}=\frac{\pi}{2}$ (to make the height at $z=0$ the node) into Eqs. (\ref{equa62}) $-$ (\ref{equa64}), the variations of the density, the longitudinal magnetic field, and the longitudinal velocity are given by, 
\begin{eqnarray}
\delta \rho = - \delta \tilde{\rho} ~\mbox{cos} (k c_T t)  ~\mbox{sin}(k z), \label{equa64.1} \\
\delta  B_{\|}= \frac {4\pi c_{s0}^2}{B_{\|}} \delta \tilde{\rho} ~\mbox{cos} (k c_T t)  ~\mbox{sin}(k z), \label{equa64.2} \\
\delta  v_{\|}=\frac { c_{s0}^2}{\rho c_T} \delta \tilde{\rho} ~\mbox{sin} (k c_T t)  ~\mbox{cos}(k z), \label{equa64.3}
\end{eqnarray}
Equations (\ref{equa64.2}) and (\ref{equa64.3}) indicate that the phase difference between the fluctuations in the magnetic field and the velocity $\phi_{B}-\phi_{v}$ is given by,
\begin{eqnarray}
\left\{
\begin{array}{l}
90^{\circ} \ {\rm for} \ n \pi  \leq kz \leq (n+\frac{1}{2}) \pi  \ ({\rm sector \ (a) \ in Fig. \ \ref{fig9}}), \label{equa64.4} \\
-90^{\circ} \ {\rm for} \ (n+\frac{1}{2}) \pi  \leq kz \leq (n+1) \pi  \ ({\rm sector \ (b) \ in Fig. \ \ref{fig9}}), 
\end{array}
\right.
\end{eqnarray}
where {$n = -1, -2, -3, ...$} . Equation (\ref{equa64.4}) indicates that the observed phase difference $\phi_{B} - \phi_{v} \sim -90^{\circ}$ is consistent with the situation that the line-forming layer is located in the sector (b).
 
Eqs. (\ref{equa64.2}) and (\ref{equa64.3}) are reduced to,
\begin{eqnarray}
\frac{|\delta B_{\|}|}{|\delta v_{\|}|}=\frac{4 \pi c_{s,0}^{2} \delta \tilde{\rho} |\sin (kz)|/B_{\|}}{c_{s,0}^{2}\delta \tilde{\rho} |\cos (kz)|/\rho c_{T}}=\frac{4 \pi \rho c_{T}|\tan (kz)|}{B_{\|}}, \label{equa65}
\end{eqnarray}
where $|\delta B_{\|}|$ and $|\delta v_{\|}|$ are amplitudes of longitudinal fluctuations in the magnetic field and the velocity. $B_{\|}$, $|\delta B_{\|}|$, and $|\delta v_{\|}|$ are related to the observables, 
\begin{eqnarray}
B_{\|}=B_{0}, \label{equa66} \\
\delta B_{\|}=\frac{\sqrt{2} \delta \Phi_{\rm los, rms}}{\overline{f}\cos{\theta}}, \label{equa67} \\
\delta v_{\|}=\frac{\sqrt{2} \delta v_{\rm los, rms}}{\cos{\theta}}. \label{equa68}
\end{eqnarray}
Since $c_{s}=\sqrt{\frac{\gamma k_{B}T}{m}}$ and $v_{A}=\frac{B_{\|}}{\sqrt{4 \pi \rho}}$, 
\begin{eqnarray}
c_{T}^{2}=\frac{c_{s}^{2}v_{A}^{2}}{c_{s}^{2}+v_{A}^{2}}=\frac{\gamma k_{B}T B_{\|}^{2}}{4 \pi \rho \gamma k_{B}T+B_{\|}^{2}m}. \label{equa69}
\end{eqnarray}
>From Eqs. (\ref{equa65}) and (\ref{equa69}), we have
\begin{eqnarray}
\Bigl( \frac{|\delta B_{\|}|}{|\delta v_{\|}|} \Bigr) ^{2}=\frac{(4 \pi \rho)^{2} \gamma k_{B}T|\tan (kz)|^{2}}{4 \pi \rho \gamma k_{B}T+B_{\|}^{2}m}. \label{equa70}
\end{eqnarray}
Equation (\ref{equa70}) leads to a second order equation for $\rho$,
\begin{eqnarray}
a_{1} \rho^{2}-a_{2} \rho -a_{3}=0, \label{equa71} \\
a_{1}=(4 \pi |\delta v_{\|}|)^{2}\gamma k_{B}T |\tan(kz)|^{2}, \label{equa72} \\ 
a_{2}=4 \pi \gamma k_{B} T |\delta B_{\|}|^{2} \label{equa73} \\
a_{3}=(B_{\|}|\delta B_{\|}|)^{2}m \label{equa74}
\end{eqnarray}
Since $\rho >0$, we can take only $\rho =\frac{a_{2}+\sqrt{a_{2}^{2}+4a_{1}a_{3}}}{2 a_{1}}$. This indicates that we can determine the mass density inside the flux tube with the additional knowledge of $\tan(kz)$ for the line-forming height. However, there are multiple solutions due to ambiguity in $\tan(kz)$. The region that we chose for the photospheric seismology (region \#02 with the assumption of the kink wave and \#05 with the assumption of the sausage wave) are both pores, whose magnetic field strength is almost the same. We assume that the parameters of the flux tube (the mass density, plasma beta, and Alfv\'en velocity) and distance between the boundary and the line formation layer derived from the analysis of the region \#05 (sausage-wave dominant) should be consistent with those derived from the analysis of the region \#02 (kink-wave dominant, section 5.5). The choice of $kz=-619^{\circ}$ for Eq. (\ref{equa40}) and $kz=-496^{\circ}$ for Eq. (\ref{equa72}) in the following exercise is based on the assumption. 

Substituting $B_{0}=1.7 \times 10^{3}$ G, $\delta \Phi_{\rm los, rms}=13.8$ G, $\delta v_{\rm los, rms}= 0.12$ km s$^{-1}$, $\overline{f}=0.81$, $\theta=29^{\circ}$, $\gamma =5/3$, $k_{B}=1.4 \times 10^{-16}$ erg K$^{-1}$, $T=1.0 \times 10^{4}$ K, and $m=1.9 \times 10^{-24}$ g, we obtain the mass density inside the flux tube $\rho = 0.8 \times 10^{-7}$g cm$^{-3}$. We then derive the values associated with the flux tube; (1) Alfv\'en speed $v_{A}=\frac{B_{0}}{\sqrt{4 \pi \rho}}$, (2) plasma beta $\beta=\frac{\rho k_{B}T/m}{B_{0}^{2}/8 \pi}$, (3) phase speed of the slow sausage mode $c_{T}$, (4) wavelength of the slow sausage mode $L=c_{T}P$, where $P$ is the observed oscillation period, and (5) distance between the boundary and the line formation layer $d=L\frac{|kz|}{360}$.  Substituting $P=294$ s, we obtain $v_{A} = 23$ km s$^{-1}$, $\beta =0.18$, $c_{T}=8.5$ km s$^{-1}$ (sound speed $c_{s}=11$ km s$^{-1})$, $L=2.5 \times 10^{3}$ km, and $d=4.3 \times 10^{3}$ km. 

The set of parameters derived here satisfy the condition that $\rho$ (or $\rho_{i}$), $\beta$, $v_{A}$, and $d$ derived here are consistent with those derived in section 5.5. The distance between the boundary and the line formation layer $d=3.6 \times 10^{3}$ km is consistent with the distance between the line-formation height and the transition region. This indeed indicates that the transition region is the reflecting layer for such waves. Note that the wavelength $L$ is much larger than the scale height $H=\frac{k_{B}T}{mg} \sim 3.9 \times 10^{2}$ km, and the effect of the gravity has to be taken into account for more rigorous treatment. 

\section{Discussions}
We have detected clear signatures of the MHD waves propagating along the magnetic flux tubes in a form of velocity, magnetic and intensity sinusoidal waves with exactly the same period. One or two strong and sharp peaks with common periods in the power spectra of the l.o.s. magnetic flux, the l.o.s. velocity, and the intensity time profiles are evident in the pores (8 peaks) and the IMSs (12 peaks). We note that only about half of the observed flux tubes have such common peaks. Periods of the peaks concentrate at around 3$-$6 minutes for pores and 4$-$9 minutes for IMSs. Phase difference between the l.o.s. magnetic flux $(\phi _{B})$, the l.o.s. velocity $(\phi _{v})$, the line core intensity $(\phi _{I,\rm core})$, and the continuum intensity $(\phi _{I,\rm cont})$ have striking concentrations at around $-90^{\circ}$ for $\phi _{B}-\phi _{v}$ and $\phi _{v}-\phi _{I,\rm core}$, around $180^{\circ}$ for $\phi _{I,\rm core}-\phi _{B}$, and around $10^{\circ}$ for $\phi _{I,\rm core}-\phi _{I,\rm cont}$ (Fig. \ref{fig5}). These fluctuations are associated with the intensity fluctuations $\frac{\delta I_{\rm cont,rms}}{\overline{I_{\rm cont}}}$ and $\frac{\delta I_{\rm core,rms}}{\overline{I_{\rm core}}}$. The amplitude of the intensity fluctuations amount to 0.1$-$ 1.0 \% of the average intensity level. Some flux tubes have a very small intensity fluctuation, and the wave mode for such flux tubes is considered to be the incompressible kink mode. On the other hand, flux tubes with higher intensity fluctuation may have the compressible sausage mode. 

The phase relation $\phi _{I}-\phi _{B} \sim 180^{\circ}$ from the observation would not be consistent with that caused by the opacity effect (e.g. Bellot Rubio et al., 2000), if the magnetic field strength decreases along the line of sight toward the observer. We propose that the longitudinal and/or transverse MHD waves propagating along the flux tube are responsible for the fluctuations. The observed phase difference $\phi _{B}-\phi _{v} \sim -90^{\circ}$ is consistent with the phase relation of the superposition of the ascending and the descending kink wave. This indicates that the ascending kink wave is substantially reflected at the chromospheric-coronal boundary. The superposed waves have the property of the standing waves. In addition to the standing kink mode, the observed phase relation between the fluctuations in the magnetic flux and the velocity is consistent with the phase relation for the superposition of the ascending and the reflected descending slow-mode sausage waves. 

So far our analysis is based on the assumption that the either the kink mode or the sausage mode is dominant in the flux tubes. Both the kink mode and the slow sausage mode may be excited in the same flux tube. Torsional waves are not discussed in this paper. The region of interest encompasses the entire magnetic flux concentrations in the spatial and temporal domain (i.e. in the case of IMSs), and we average the physical parameters inside the ROI. Thus, we are probably unable to detect the torsional Alfv\'en waves, even if they exist, because the perturbation of the magnetic flux and  the velocity is averaged over the whole flux tube, and are canceled out. 

We derive the various physical parameters of the flux tubes only from the observed period and the amplitudes of magnetic and velocity oscillations. Such parameters include (1) mass density inside and outside the flux tube, (2) plasma $\beta$ inside the flux tube, (3) Alfv\'en speed inside the flux tube, (4) phase speed, (5) wavelength, (6) distance between the boundary and the line formation layer, and (7) propagation time of fast magneto-acoustic wave across the flux tube. In the examples presented in this paper, we choose similar sets of $kz$ as defined in sections 5.5 and 6.3 for both cases (the kink-wave dominant case and the sausage-wave dominant case) such that the derived physical parameters of the flux tubes coincide. The choice determines the distance $d$ between the boundary (node) and the line formation layer. The derived mass density outside the flux tube is consistent with that of the standard solar model in the case of the kink wave. Note that we can not derive the mass density outside the flux tube in the case of the slow sausage mode, because the flux tube is not in the pressure equilibrium. This exercise demonstrates that the seismology of magnetic flux tubes is possible with the observations of the oscillation period and amplitudes for various photospheric and chromosheric lines, and may open a new channel for the diagnostics of the magnetic flux tubes.  

Magnetohydrodymanic waves are believed to play a vital role in the acceleration and heating of the fast solar wind. However, it has been thought that the Alfv\'en speed rapidly increases with height due to the rapid decrease in the plasma density, resulting in significant reflection at the chromosphere-corona boundary. We indeed show that this may be the case in this paper: the upward propagating kink and/or sausage waves must be significantly reflected back above the line formation layer. Deviation in the phase difference between the magnetic and velocity fluctuations from $-90^{\circ}$ as seen in Fig. 6 may indicate residual waves propagating to the corona. Indeed, the upward Poynting flux above the reflecting layer is estimated to be $2.7 \times 10^{6}$ erg cm$^{-2}$ s$^{-1}$ in one case (kink wave), and is by no means negligible flux in terms of heating and acceleration of the upper atmosphere.  

Tsuneta et al (2008b) conjectures that a rapid decrease in the magnetic field strength associated with the rapidly expanding flux tube near the chromosphere-corona boundary for the polar kG patches reduces the vertical change in Alfv\'en speed, and the Alfv\'enic cutoff frequency be lower in the polar flux tubes. Magnetohydrodymanic waves generated in the photosphere may be more efficiently propagated to the corona through such fanning-out flux tubes with large expansion factor observed in the polar coronal holes. On the other hand, the observations presented here suggest significant reflected waves. It would therefore be interesting to see whether the reflectivity of the magnetohydrodymanic waves depends on the locations or environment e.g. coronal holes vs the quiet Sun.  

Two interpretations addressed here (kink and sausage MHD modes) cannot be distinguished in the present study. The flux tubes that we analyzed are located with angular distance of 25$-$49$^{\circ}$ from the Sun center for high sensitivity magnetic observations. It is important to compare the wave properties for the flux tubes located further away from the Sun center with those of the flux tubes around the disk center to separate individual modes of waves (Norton et al. 2001). These topics will be addressed in our subsequent paper.

\section*{Acknowledgments}
We gratefully thank M. Ryutova and O. Steiner for the fruitful discussions on the sausage mode and the opacity effect. M. Ryutova helped us to theoretically formulate the properties of the sausage mode in section 6.1. We acknowledge encouragements from T. Yokoyama, E. Priest, B. Roberts, R. Erd\'{e}lyi, E. Khomenko, N.Yokoi and M. Velli. 

Hinode is a Japanese mission developed and launched by ISAS/JAXA, collaborating with NAOJ as a domestic partner, NASA and STFC (UK) as international partners. Scientific operation of the Hinode mission is conducted by the Hinode science team organized at ISAS/JAXA. This team mainly consists of scientists from institutes in the partner countries. Support for the post-launch operation is provided by JAXA and NAOJ (Japan), STFC (U.K.), NASA, ESA, and NSC (Norway). This work was carried out at the NAOJ Hinode Science Center, which is supported by the Grant-in-Aid for Creative Scientific Research "The Basic Study of Space Weather Prediction" from MEXT, Japan (Head Investigator: K. Shibata), generous donations from Sun Microsystems, and NAOJ internal funding. 

\section*{Appendix}
\begin{figure}
\epsscale{0.8}
\plotone{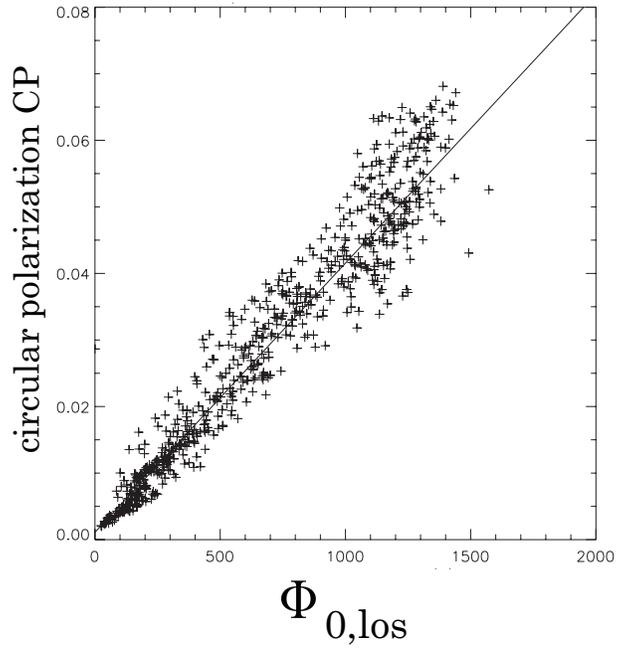}
\caption{
Scatter plot indicating the relation between the l.o.s. magnetic flux $\Phi_{0,\rm los}$ (see text) and the circular polarization $CP$ as defined in equation (\ref{equa1}). Solid line indicates a linear regression line.
}
\label{fig10}
\end{figure}

We here determine the conversion coefficient $\lambda$ to convert the circular polarization CP to the l.o.s. magnetic flux in equation (\ref{equa4}). Fig. \ref{fig10} shows the scatter plot for the circular polarization derived by equation (\ref{equa1}) with the l.o.s. magnetic flux. The l.o.s. magnetic flux $\Phi_{0,\rm los}=B_{0,\rm los} f$ is determined from the line-of-sight magnetic field strength $B_{0,los}$ and the filling factor $f$, both obtained with the Milne-Eddington inversion. The data used here is the plage region \#05 (Table \ref{table1}) taken at 19:15 UT on 2007 Feb. 3. We notice a good linear correlation between the two quantities. The linear regression is given by
\begin{eqnarray} 
CP=(4.16 \times 10^{-5}) \Phi_{0,\rm los}+0.0016. \label{equa75}
\end{eqnarray}
The correlation coefficient is 0.96. We use the conversion coefficient $\lambda =4.16 \times 10^{-5}$ G$^{-1}$ for the analysis presented in this paper. 

\end{document}